\documentclass[10pt,a4paper]{article}

\usepackage[a4paper,text={16.8cm,22.4cm}]{geometry}
\usepackage{amsmath,amsfonts,slashed,amssymb,tikz,bm,psfrag,graphicx,color,dsfont}
\usepackage{multicol}
\usepackage{float}
\usepackage{slashed}
\usepackage{euscript}
\usepackage[normalem]{ulem}
\RequirePackage[sort&compress,square,comma,numbers]{natbib}
\allowdisplaybreaks
\renewcommand{\arraystretch}{1.2}
\usepackage[utf8]{inputenc}
\usepackage[T1]{fontenc}
\usepackage{graphicx}
\usepackage{amsmath}
\usepackage{amsfonts}      
\usepackage{empheq}       
\usepackage{hyperref}     
\usepackage{euscript}
\usepackage{multirow}
\usepackage{simplewick}
\usepackage{xcolor}
\usepackage{slashed}
\usepackage{rotating} 
\usepackage{pgfplots}
\pgfplotsset{compat=1.18}
\usepackage{filecontents}
\usepackage[utf8]{inputenc}
\usepackage{placeins}

\hypersetup{
  colorlinks   = true,    
  urlcolor     = blue,    
  linkcolor    = blue,    
  citecolor    = red      
}

\usepackage{ifpdf}

\begin{document}

\begin{titlepage}

\begin{flushright}
\normalsize
\end{flushright}

\vspace{0.1cm}
\begin{center}
\Large\bf
$B \to D$ form factors from light-cone sum rules
\end{center}

\vspace{0.5cm}
\begin{center}
{\bf  Yang Qu$^{a}$, Jin-Peng Jiang$^{a}$, Yue-Long Shen$^{a,b,c}$ \footnote{ shenylmeteor@ouc.edu.cn}
}\\
\vspace{0.7cm}
{\sl ${}^a$ College of Physics and Optoelectronic Engineering,\\ Ocean University of China, Qingdao 266100, China}\\
{\sl${}^b$Engineering Research Center of Advanced Marine Physical Instruments and Equipment,\\ Ministry of Education, Qingdao 266100, China}\\
{\sl${}^c$Qingdao Key Laboratory of Optics and Optoelectronics, Qingdao 266100, China}
\end{center}

\vspace{0.2cm}
\begin{abstract}

We present a light-cone sum rule (LCSR) analysis of the $B\to D$ transition
form factors $f_+^{BD}(q^2)$ and $f_0^{BD}(q^2)$ in the large-recoil region,
including the next-to-leading-power corrections at tree level. The
non-perturbative input is provided by the leading-twist $D$-meson
distribution amplitude determined from lattice QCD in the heavy-quark large-momentum effective theory (HQLaMET) framework,
which enters together with the twist-three and twist-four distribution
amplitudes of the $D$ meson fixed by the QCD equations of motion. A combined fit to the LCSR
results at $q^2\lesssim2$~GeV$^2$ and to the FNAL/MILC and HPQCD lattice data
at $q^2\gtrsim8.5$~GeV$^2$, performed in the model-independent Boyd-Grinstein-Lebed (BGL)
parametrisation, allows us to obtain the form factors over the whole physical
range and to extract $|V_{cb}|$ from the semileptonic $B\to D\ell\nu$ decay as
well as the lepton-flavour universality ratio $R(D)$.

\end{abstract}

\vfil

\end{titlepage}

\tableofcontents

\newpage

\section{Introduction}

The determination of the Cabibbo--Kobayashi--Maskawa (CKM) matrix elements is
one of the primary goals of heavy-flavour physics. To date, a tension of
$2$--$3\sigma$ still persists between the exclusive and inclusive
determinations of $|V_{cb}|$. Considerable effort has been devoted to
identifying the origin of this long-standing puzzle, both within the Standard
Model (SM) and in scenarios beyond it~\cite{Belle-II:2018jsg,Gambino:2020jvv,Boyle:2022uba,USQCD:2022mmc}.
The semileptonic decays $\bar{B}_{(s)} \to D_{(s)}^{(*)} \bar{\ell}\nu_{\ell}$
are essential for a precise extraction of $|V_{cb}|$. Moreover, for the
lepton-flavour-universality (LFU) ratio
$\mathcal{R}(D^{*})=\mathcal{B}(B\to D^{(\ast)}\tau \nu_\tau)/\mathcal{B}(B\to
D^{(\ast)}\ell \nu_\ell)$, the state-of-the-art lattice predictions exhibit a
notable tension with the experimental measurements, reaching a combined
significance of about $3\sigma$. This discrepancy has attracted extensive
attention in recent years, since it may signal physics beyond the
SM~\cite{Bernlochner:2021vlv,Bernlochner:2016bci,Bernlochner:2018kxh,Cohen:2018vhw,Das:2021lws,Gubernari:2022hrq,
	Bernlochner:2022hyz,Patnaik:2022moy,Patnaik:2023ins,Penalva:2023snz}.

The $\bar{B}_{(s)} \to D_{(s)}^{(*)}$ form factors are the fundamental
non-perturbative inputs of the semileptonic decays
$\bar{B}_{(s)} \to D_{(s)}^{(*)} \bar{\ell}\nu_{\ell}$, and they play a
central role in the extraction of $|V_{cb}|$ and in the prediction of
$\mathcal{R}(D^{(*)})$. In the low-recoil region they can be studied in a
model-independent way in heavy-quark effective theory (HQET), which organises
them as an expansion in powers of $\Lambda_{\rm QCD}/m_{b,\,c}$ in terms of
the non-perturbative Isgur--Wise (IW) functions. Lattice QCD also provides
reliable determinations of these form factors at zero recoil, and substantial
progress has been achieved in the lattice study of semileptonic $B$-meson
decays in recent years. For the $B\to D\ell\bar{\nu}$ transition, the first
unquenched lattice determination of the corresponding hadronic form factors
was reported by the Fermilab--MILC collaboration in
Ref.~\cite{FermilabLattice:2015ilb}; subsequently, the HPQCD collaboration
performed full-$q^2$ unquenched calculations of the form factors $f_+(q^2)$
and $f_0(q^2)$~\cite{Na:2015kha}. Lattice results, however, have to be
extrapolated over the full physical kinematic range by means of
phenomenological models
\cite{Boyd:1995cf,Boyd:1995sq,Boyd:1997kz,Caprini:1996vw,Caprini:1997mu,Bourrely:2008za},
and such an extrapolation inevitably introduces sizeable uncertainties in the large-recoil
region.

By contrast, the light-cone sum rule (LCSR) approach, which rests on
parton-hadron duality and on the light-cone operator product expansion
(OPE), provides a suitable framework for computing heavy-to-heavy transition
form factors directly in the large-recoil region
\cite{DeFazio:2005dx,DeFazio:2007hw,Khodjamirian:2005ea,Khodjamirian:2006st,Wang:2017jow,Gao:2021sav,Cui:2022zwm,Cui:2023jiw,Khodjamirian:2023wol,Gao:2024vql}.
It has already yielded extensive predictions for the exclusive $B\to D$ form
factors involving an energetic charmed meson. Within this framework, QCD
factorisation of the leading-power contribution to the vacuum-to-$B$
correlation function, built with the $D$-meson interpolating current, has
been worked out systematically at tree level and at one-loop level
\cite{Wang:2017jow,Gao:2021sav}. Subleading power corrections, induced by
higher-twist light-cone distribution amplitudes (LCDAs) and by the
$\Lambda_{\rm QCD}/m_b$ corrections of the heavy-quark expansion in HQET, are
also properly included. Combining these power-suppressed contributions with
the large-recoil LCSR results and with the lattice data, combined fits of the
$B\to D$ form factors have been performed, which significantly improve the
previous determinations of the LFU ratios $\mathcal{R}(D^{(*)})$
\cite{Cui:2023jiw}. At large recoil the $B\to D$ form factors have also been
computed in the PQCD approach based on $k_T$ factorisation
\cite{Kurimoto:2002sb}.

The $B\to D$ form factors can alternatively be obtained from LCSR with a
vacuum-to-$D$-meson correlation function interpolated by a $B$-meson state
\cite{Zuo:2006dk,Fu:2013wqa,Zhang:2017rwz,Wu:2025kdc}. In this approach the
infrared dynamics is encoded in the $D$-meson LCDAs. Since
$m_Q \gg \Lambda_{\rm QCD}$, two classes of heavy-meson LCDAs are required:
the QCD LCDAs, relevant when the momentum transfer is much larger than the
heavy-quark mass, and the HQET LCDAs, defined in the heavy-quark limit. Until
recently the $D$-meson QCD LCDAs were poorly constrained. The large-momentum
effective theory (LaMET) \cite{Ji:2013dva,Ji:2014gla,Ji:2020ect}   now provides a systematic framework for determining
matrix elements of non-local operators with light-cone separation: rather
than relying on a local OPE, it relates Euclidean equal-time correlators of a
boosted hadron to their Minkowski counterparts through factorisation and
perturbative matching. This framework has been employed to determine the LCDAs of light mesons, including the pion and kaon, to high precision \cite{LatticeParton:2022zqc,Liu:2018tox}.

In this work, we employ the recently updated $D$-meson LCDAs from lattice simulations~\cite{LPC:2026vyv,Han:2024fkr} to study the $B\to D$ form factors within the framework of $D$-meson light-cone sum rules. For the leading-power contributions, we evaluate the hard kernels through QCD-to-SCET matching and perform the resummation at next-to-leading-logarithmic (NLL) accuracy. Higher-twist LCDAs remain inaccessible to lattice QCD and are estimated only phenomenologically. To improve the theoretical precision, we include the two-particle twist-3 and twist-4 LCDAs, which are related to their twist-2 counterparts by the equations of motion \cite{Ball:2006wn}, while neglecting the suppressed three-particle contributions; the finite charm-quark mass and $D$-meson mass corrections are retained throughout. We also provide predictions for the branching ratios of $B\to D\ell\nu_{\ell}$, for which independent fits with the BGL parametrisation of the form factors are indispensable~\cite{Belle:2018ezy}. We then extract the CKM matrix element $|V_{cb}|$ from the semileptonic $B\to D\ell\nu$ decay as well as the lepton-flavour universality ratio $R(D)$.

The paper is organised as follows. In Sec.~2, we derive the light-cone sum rules for the $B\to D$ form factors in terms of the $D$-meson LCDAs, including the leading-twist contributions and the power corrections to the correlation function. In Sec.~3, we present the numerical results for the form factors and for the branching ratios of $B\to D\ell\nu$, together with the extraction of $|V_{cb}|$ and the determination of $R(D)$. In Sec.~4, we summarise our findings.

\section{Light-cone sum rules for $B \to D$ form factors }

In this paper, we concentrate on the semileptonic decays $B \to D \ell \bar{\nu}_\ell$ with charged leptons $\ell = e,\,\mu$, for which only the form factors associated with vector and axial-vector currents contribute. The $B \to D$ transition form factors $f_+(q^2)$ and $f^-(q^2)$ are conventionally defined via the matrix element
\begin{equation}
	\langle D(p) | \bar{c}\gamma_\mu b |\bar B(p+q) \rangle
	= 2f^+(q^2)\,p_\mu + \big[f^+(q^2)+f^-(q^2)\big]\,q_\mu,
\end{equation}
where $p$ denotes the four-momentum of the $D$ meson, and $q$ stands for the momentum transfer between the initial and final hadronic states. An alternative parametrization of the $B \to D$ decay amplitude employs two form factors $F^+(q^2)$ and $F^0(q^2)$, arising from the full Lorentz decomposition of the corresponding hadronic matrix element:
\begin{eqnarray}
\langle D(p) | \bar{c}\gamma_\mu b | \bar{B}(p+q) \rangle
= F^+(q^2) \left[ 2p_\mu + \left( 1 - \frac{m_B^2 - m_D^2}{q^2} \right) q_\mu \right]
+ F^0(q^2) \frac{m_B^2 - m_D^2}{q^2}\,q_\mu.
\end{eqnarray}
These two parametrizations are related via the identities below.
\begin{eqnarray}
	 F^+(q^2) &=&f^+(q^2) \nonumber \\
	  F^0(q^2) &=&f^+(q^2)+{q^2\over m_B^2 - m_D^2}f^-(q^2),
\end{eqnarray}
Obviously, we have $F^+(0) = F^0(0)$.

To determine the form factors defined above, the correlation function is adopted as follows
\begin{eqnarray}
	\Pi_\mu(p,q) &= i \int d^4 x e^{i q \cdot x} \langle D(p) | T \left\{ \bar{c}(x) \gamma_\mu  b(x), m_b\bar{b}(0) i\gamma_5 q(0) \right\} | 0 \rangle 
\end{eqnarray}
 In the time-like region ($q^2>0$), by inserting a complete set of intermediate hadronic states into the correlator, one can obtain its hadronic representation by isolating the pole term of the lowest state of the $B$ meson. By further using the definition of the form factors and the $B$ meson decay constant 
\begin{equation}
	\langle B | m_b\bar{b}i\gamma_5 q | 0 \rangle = m_{B_q}^2 f_{B_q}, 
\end{equation}
where $m_{B_q}$ is the $B_q$-meson mass and $m_b$ is the $b$-quark mass, the hadronic representation for the correlator  reads
\begin{equation}
		\Pi_\mu^{\text{had}}(p,q)= \frac{\{2f^+(q^2)\,p_\mu + [f^+(q^2)+f^-(q^2)]q_\mu\} m_{B}^2 f_{B}}{ m_{B}^2 - P^2} + \int_{s_0^{B}}^\infty \frac{\rho^{\mathrm{h}}(s)n_\mu +\tilde \rho^{\mathrm{h}}(s)\bar n_\mu }{s - P^2} ds,
\end{equation}
where $s_0^{B}$ is the threshold parameter, $\rho^{\text{h}}(s)$ and $\tilde \rho^{\text{h}}(s)$  are the spectral densities, $P=p+q$ denoting the momentum of the interpolating current of the $B$ meson. To facilitate the extraction of the leading-power form factor, the hadronic correlation function can be expressed as
\begin{eqnarray}
	\Pi_\mu^{\text{had}}(p,q)&=& \frac{ m_{B}^3 f_{B}}{ 2(m_{B}^2 - P^2)}\bigg \{\left[\left(1+{\bar n\cdot p\over m_B}\right)f^+(q^2)+\left(1-{\bar n\cdot p\over m_B}\right)f^-(q^2)\right] n_\mu \nonumber \\ &+&\left[\left(1+{m_D^2\over m_B\bar n\cdot p}\right)f^+(q^2)+\left(1-{m_D^2\over m_B\bar n\cdot p}\right)f^-(q^2)\right] \bar n_\mu\bigg\}\nonumber \\
	&+ & \int_{s_0^{B}}^\infty \frac{\rho^{\mathrm{h}}(s)n_\mu +\tilde \rho^{\mathrm{h}}(s)\bar n_\mu }{s -P^2} ds,
\end{eqnarray}
On the other hand, the correlator in the spacelike region can be evaluated via the operator product expansion (OPE).  At the partonic level, the correlator can be parametrised as 
\begin{align}
	\Pi_{\mu}^{\mathrm{par}}(p,q) &= 
	\Pi(P^2,q^2)\,{n}_\mu
	+\tilde\Pi(P^2,q^2)\,\bar n_\mu,
\end{align}

To extract the $B\to D$ form factors, we take advantage of the dispersion relation and express the partonic level representation of the correlation function as
\begin{align}
	\Pi(\tilde \Pi)(p,q)
	= \int_{0}^{\infty} \frac{ds}{s- P^2 -i0}
	\frac{1}{\pi}\mathrm{Im}\Pi(\tilde \Pi)(s,q^2). 
\end{align}

In the LCSR approach,  the partonic representation $\Pi_{\mu}^{\mathrm{par}}$ derived with the perturbative factorization approach should be matched to the  physical one $\Pi_{\mu}^{\mathrm{had}}$ obtained with the hadronic dispersion relation. The quark-hadron duality plays an essential role in the matching procedure; in our case it takes the following form 
\begin{align}
	\int_{s_0^{B}}^\infty \frac{\rho^{\mathrm{h}}(s)(\tilde \rho^{\mathrm{h}}(s)) }{s -P^2} ds
	= 	 \frac{1}{\pi}\int_{S_0^{\rm th}}^{\infty} \frac{ds}{s- P^2 } \mathrm{Im}\Pi(\tilde \Pi)(s,q^2). 
\end{align}
After subtracting the continuum and completing the matching, we can compute the physical quantities in the large-recoil region. However, when analysing the LCSR numerical values, one must be cautious as parton-hadron duality can introduce systematic errors that need to be carefully controlled. We introduce the Borel transformation to weaken the irrelevant contributions of the continuum spectrum and finally derive the form factor formula.
\begin{eqnarray}
\left[\left(1+{\bar n\cdot p\over m_B}\right)f^+(q^2)+\left(1-{\bar n\cdot p\over m_B}\right)f^-(q^2)\right] &= &	\frac{2}{\pi m_{B}^3 f_{B}}\int^{s_0^{\rm th}}_{m_b^2} ds{\rm exp}\left[-{s-m_B^2\over M^2}\right] \mathrm{Im}\Pi(s,q^2)\nonumber\\ 
\left[\left(1+{m_D^2\over m_B\bar n\cdot p}\right)f^+(q^2)+\left(1-{m_D^2\over m_B\bar n\cdot p}\right)f^-(q^2)\right] &= &\frac{2}{\pi m_{B}^3 f_{B}}\int^{s_0^{\rm th}}_{m_b^2} ds{\rm exp}\left[-{s-m_B^2\over M^2}\right]\mathrm{Im}\tilde \Pi(s,q^2) ,
\label{sumrulemain}\end{eqnarray}
Here we have neglected the light-quark mass. In the following we compute the correlation function with the QCD factorization approach. 
\subsection{Leading power contribution to the correlation function}
\subsubsection{Contribution at tree level}
 \begin{figure}[h]
	\begin{center}
		\vspace{0.3cm}\hskip0cm
		\includegraphics[width=0.3\textwidth]{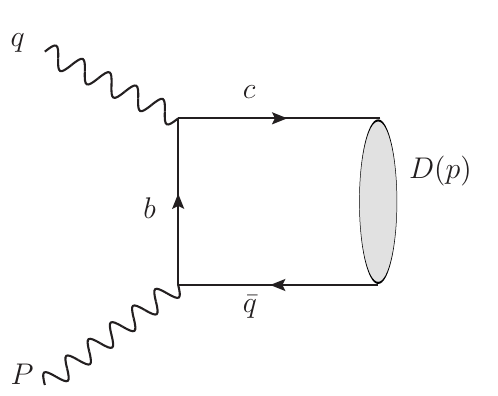}
	\end{center}
	\caption{ Leading order diagrams of  the correlation function. }
	\label{fig:LO}
\end{figure}
We calculate the correlation function in momentum space.  The corresponding Feynman diagrams are depicted in Figure \ref{fig:LO}.
At leading order, the partonic representation of the correlation function reads
\begin{eqnarray}
	\Pi_{\mu}^{{\rm LP}} 
	&=& -{im_b^2\over  uP^2-m_b^2+\bar uq^2}\bar u(k) {\not\! \bar n \over 2} \gamma_5 v(p-k) n_\mu\,
	\nonumber \\
	&\equiv&- {im_b^2\over vP^2-m_b^2+\bar vq^2}\ast \langle O_1(v,u)\rangle^{(0)} n_\mu\,.
\end{eqnarray}
Here, $\langle O_1\rangle^{(0)}$ 
denotes the tree-level partonic matrix element
of the operator $O_1$ in the framework of  soft-collinear effective theory (SCET) \cite{Bauer:2001rh}
\begin{eqnarray}
	\langle O_1(u, v)  \rangle ^{(0)} \equiv
	\langle q(u\, p) \, \bar q (\bar u \, p) | O_1^{(0)} (v) | 0 \rangle
	=\bar u(u \, p)  \,{\not\! \bar n \over 2} \gamma_5 \,v(\bar u \, p) \,\, \delta(u - v)  \,,
\end{eqnarray}
and  the convolution integration is represented by an asterisk.
The manifest definition of the SCET operator $O_1$ in momentum space is given by
\begin{eqnarray}
	O_1(v)= {\bar n \cdot p \over 2 \pi}  \, \int d \tau  \, e^{i \, v \, \tau \, \bar n \cdot p} \,\,
	\bar \xi (\tau \bar n) \, W_c(\tau \bar n,0) \, {\not\! \bar n \over 2} \  \gamma_5 \,  \xi(0) \,,
\end{eqnarray}
with the collinear Wilson line
\begin{eqnarray}
	W_c(\tau \bar n,0) = {\rm P} \, \left \{ {\rm  Exp} \left [   i \, g_s \,
	\int_{0}^{\tau} \, d \lambda \,  \bar n  \cdot A_{c}(\lambda \, \bar n) \right ]  \right \}.
\end{eqnarray}

We now employ the operator matching equation
\begin{eqnarray}
\Pi_{\mu}^{{\rm LP},(0)}  =- \, \, T_{1}^{(0)} (v)   \, \ast \langle O_1(u,v) \rangle ^{(0)} \,  n_\mu,
	\label{matching equation}
\end{eqnarray}
and expand all quantities at tree level, yielding
\begin{eqnarray}
	T^{(0)}_{1}(u)= {m_b^2 \over uP^2-m_b^2+\bar uq^2}\equiv{1\over -\bar \rho}      \,,  \qquad
	\end{eqnarray}
	In the expressions above, we adopt the notation
	$r_1=q^2/m_b^2$, $r_2=P^2/m_b^2$,  $\rho=r_1+u(r_2-r_1)$. 
Utilizing the definition of the leading-twist $D$-meson LCDA on the light-cone
\begin{eqnarray}
	\langle  D(p) |\bar \xi (y) \, W_c(y, 0) \, \gamma_{\mu} \, \gamma_5 \, \xi(0) |  0 \rangle
	= -i \, f_{D} \, p_{\mu}  \, \int_0^1 \, d u \, e^{i\, u \, p \cdot y}  \, \phi_{D} (u, \mu)
	+ {\cal O}(y^2) \,,
\end{eqnarray}
it is straightforward to derive the tree-level factorization formula of the correlation function
\begin{eqnarray}
&&	\Pi^{{\rm LP},(0)}  =-{1\over 2}f_D\bar n\cdot p \ \int _0^1 du 	T^{(0)}_{1}(u,\mu)  \phi_{D} (u, \mu)\, ,\nonumber  \\
&&	\tilde\Pi^{{\rm LP},(0)}  =0\,	
\end{eqnarray}
We express it as a dispersion integral and integrate over the variable $s$, which gives
\begin{eqnarray}
	\Pi^{{\rm LP},(0)}(P^2,q^2)  ={1\over 2}f_D\bar n\cdot p m_b^2 \int _0^{1}  {du\over u} {\phi_{D} (u, \mu)\over s(u)-P^2}  \, ,
\end{eqnarray}
where 
\begin{eqnarray}
	s(u)={m_b^2-\bar uq^2\over u}, \qquad 
\end{eqnarray}
At leading power, we neglect the terms proportional to $m_D^2/m_B^2$. Taking advantage of quark-hadron duality and performing the Borel transformation with respect to $P^2$, we arrive at  
\begin{equation}
	f^{+,(0)}_{LP}(q^2)=\frac{f_D m_b^2}{2f_Bm_B^2} \int _{u_0}^{1}  {du\over u} {\phi_{D} (u, \mu)} e^{(m_B^2-s(u))/M^2} , \label{srlo}
\end{equation}
with $u_0=(m_b^2-q^2)/(s_0-q^2)$.   Meanwhile,  $f^{-,(0)}_{LP}(q^2)+f^{+,(0)}_{LP}(q^2)$ vanishes at leading power.
\subsubsection{Contribution at one-loop level }
\begin{figure}[h]
	\begin{center}
		\vspace{0.3cm}\hskip0cm
		\includegraphics[width=0.45\textwidth]{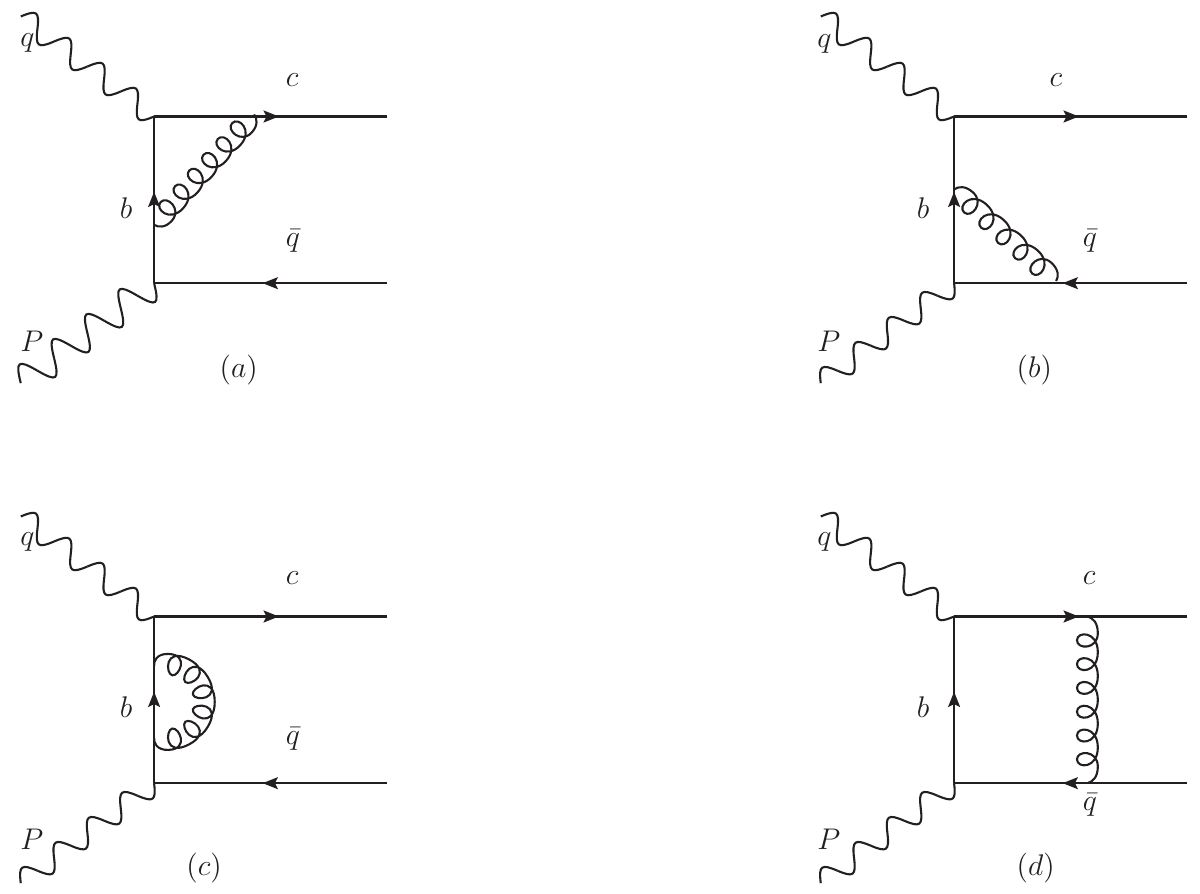}
	\end{center}
	\caption{Next-to-leading-order diagrams of the correlation function.}
	\label{fig:NLO}
\end{figure}

At NLO,  the factorization formulae for the correlation functions can be written as
\begin{eqnarray}
	&&	\Pi_{\rm LP}^{(1)} =\frac{\alpha_sC_F}{4\pi}f_D\bar n\cdot p  \int _0^1 du 	T^{(1)}_{1}(u,\mu)  \phi_{D} (u, \mu)\, ,\nonumber  \\
	&&	\tilde\Pi_{\rm LP}^{(1)}  =\frac{\alpha_sC_F}{4\pi}f_D\bar n\cdot p  \int _0^1 du 	 T^{(1)}_{2}(u,\mu)  \phi_{D} (u, \mu)\,\,	
	\label{NLO}
\end{eqnarray}
To obtain the hard functions, we employ the master formulae as follows 
\begin{eqnarray}
	&&\langle O_1(u, v)  \rangle ^{(0)} \otimes T^{(1)}_{1}(u,\mu)= \Pi^{(1)}(P,q)+Z_\Pi \Pi^{(0)}(P,q)- \langle O_1(u, v)  \rangle ^{(1)} \otimes T^{(0)}_{1}(u,\mu),\nonumber  \\
	&&\langle O_1(u, v)  \rangle ^{(0)} \otimes T^{(1)}_{2}(u,\mu)= \tilde \Pi^{(1)}(P,q).
\end{eqnarray}
where $\Pi^{(1)}(P,q)$ is obtained by evaluating the Feynman diagrams shown in Fig.~\ref{fig:NLO}. Within this calculation, both ultraviolet (UV) and infrared (IR) divergences are regularised via dimensional regularisation. A well-known subtlety for QCD corrections to the hard function in this framework concerns the definition of the Dirac matrix $\gamma_5$ in the complex $D$-dimensional space, and multiple prescriptions for treating $\gamma_5$ have been developed to fulfill the requirements of precision QCD computations under various circumstances. In the present work, we employ the naive dimensional regularization (NDR) scheme for $\gamma_5$, consistent with the scheme adopted for computing the NLO evolution kernel of the leading-twist distribution amplitude of the $D$-meson.

$Z_\Pi$ denotes the operator renormalization constant. We remark that both the transition current $\bar c \gamma_\mu b(x)$ and the interpolating current for the $B$-meson appearing in the correlation function are physical and do not require operator renormalization, so that $Z_\Pi = 1$.  The matrix element $\langle O_1(u, v) \rangle^{(1)}$ contains both UV and IR divergences, which cancel against each other within dimensional regularization. The collinear divergence can be isolated, and it cancels the corresponding divergence present in $\Pi^{(1)}(P,q)$, as illustrated in
\begin{eqnarray}
	\langle O_1(u, v)  \rangle ^{(1)} = -\frac{\alpha_s C_F}{4\pi} \frac{2}{\tilde \epsilon} \left[ \frac{\bar v}{\bar{u}} \frac{v + \bar{u}}{v - u} \theta(v - u) + \frac{v}{u} \frac{u + \bar v}{u - v} \theta(u - v) \right],\qquad \frac{1}{\tilde\epsilon}=\frac{1}{\epsilon}-\gamma_E+\ln 4\pi,
\end{eqnarray}
As expected, it is proportional to the evolution kernel of the QCD distribution amplitude of the $D$ meson, $V_0(u,v)$, which satisfies
\begin{eqnarray}
	\frac{d}{d\ln \mu}\phi_{D}(u,\mu)=\int_0^1 dv V_0(u,v) \phi_D(v,\mu).
\end{eqnarray}
Since $\tilde \Pi(q^2)$ vanishes at leading order,    it must be finite at one-loop level.  After explicit  calculation and subtraction of divergence,  the  hard kernels $T_{1,2}^{(1)}$ at $\mathcal {O} (\alpha_s)$ are obtained as
\begin{eqnarray}
		\label{hk1}
	T_1^{(1)}
	&&= \left(  \frac{\bar r_2}{(r_2-r_1)^2 u}-\frac{1}{\bar \rho}\right) G(r_1)
	+ \left( \frac{\bar r_1}{\bar u(r_2-r_1)^2 }-\frac{1}{\bar \rho}\right) G(r_2) \nonumber \\ \nonumber 
	&&\quad - \left( \frac{\bar r_2}{(r_2-r_1)^2 u} + \frac{\bar r_1}{(r_2-r_1)^2 \bar u}-\frac{2}{\bar \rho}\right) G(\rho)\nonumber + \frac{1}{r_2}\left( \frac{\bar r_2}{\bar \rho} + \frac{\bar r_2}{(r_2-r_1)\bar u} \right)\ln\bar r_2 \\
	&&\quad + \frac{1}{r_2}\left( \frac{r_2-2}{2\rho} - \frac{r_2}{2\rho^2} - \frac{\bar r_2}{(r_2-r_1)\bar u} \right)\ln\bar \rho \nonumber  + \frac{\rho+1}{2\bar \rho^2}\left( 3\ln\left(\frac{m_b^2}{\mu^2}\right) - \frac{3\rho+1}{\rho} \right)\\
	&&\quad+ \frac{1}{2}{\bar n \cdot q\over \bar n\cdot p} \bigg [  \frac{ r_1 r_2 -r_1^2+ \bar r_1(r_2-r_1)\ln\bar r_1}{r_1^2\bar\rho}
	+ \frac{\bar r_1(r_1+r_2)\ln\bar r_1}{r_1^2(r_2-r_1)u} + \frac{2\bar r_2\ln\bar r_2}{r_2\bar u(r_2-r_1)}
	\\\nonumber 
	&&\quad - \frac{\bar \rho (r_2+\rho)\ln\bar \rho }{(r_2-r_1)u\bar u\rho^2}
	+ \frac{r_2-r_1}{r_1\rho}  \bigg ], 
\end{eqnarray}
and 
\begin{eqnarray}
	\label{hk2}
	T_2^{(1)}
	&&=\frac{1}{2}{ n \cdot q\over \bar n\cdot p}\bigg [  \frac{ r_1 r_2 -r_1^2+ \bar r_1(r_2-r_1)\ln\bar r_1}{r_1^2\bar\rho}
	+ \frac{\bar r_1(r_1+r_2)\ln\bar r_1}{r_1^2(r_2-r_1)u} + \frac{2\bar r_2\ln\bar r_2}{r_2\bar u(r_2-r_1)}
	\\\nonumber 
	&&\quad - \frac{\bar \rho (r_2+\rho)\ln\bar \rho }{(r_2-r_1)u\bar u\rho^2}
	+ \frac{r_2-r_1}{r_1\rho}  \bigg ], 
	\end{eqnarray}
where we have used the notation $\bar x\equiv 1-x$.  The logarithmic combination $G(x)$ is defined by
\begin{eqnarray}
	G(x)=\mathrm{Li}_2(x)+\ln^2(1-x)+\ln(1-x)\left(\ln\frac{m_b^2}{\mu^2}-1\right),
\end{eqnarray} 
where $\mathrm{Li}_2(x)$ denotes the dilogarithm function.   We can readily verify that our result agrees with the leading-power hard kernel derived in \cite{Khodjamirian:1997ub,Duplancic:2008ix}, where the \(B\to\pi\) form factors were computed within the pion light-cone sum rule framework. Since the charm-quark mass is neglected at leading power, our predictions are expected to reproduce the same hard kernel as that for the \(B\to\pi\) transition.  In the factorization formula (\ref{NLO}), both the hard kernel and the LCDAs of the $D$ meson depend on the factorization scale $\mu$.  If we set  $\mu$  to be a hard scale, there must exist large  logarithms in the LCDAs.  In order to perform the leading-logarithmic QCD resummation, it is convenient to adopt the Gegenbauer expansion of the leading-twist LCDA of the $D$ meson
\begin{equation}
	\phi_D(x,\mu)=6 x \bar{x}\sum_{n=0}^{\infty} a_n(\mu)\, C_n^{3/2}(2x-1),
	\label{eq:pion_da_gegenbauer}
\end{equation}
where the LO coefficient $a_0(\mu)=1$ is renormalisation invariant.  The exact solution to the two-loop RG equation can be written as \cite{Mueller:1993hg,Mueller:1994cn}
\begin{equation}
	a_n(\mu)=E_{V,n}^{\text{NLO}}(\mu,\mu_0)\,a_n(\mu_0)
	+\frac{\alpha_s(\mu)}{4\pi}\sum_{k=0}^{n-2}E_{V,n}^{\text{LO}}(\mu,\mu_0)\,d_{V,n}^k(\mu,\mu_0)\,a_k(\mu_0),
	\label{eq:an_sol}
\end{equation}
where both $n$ and $k$ are non-negative  integers and the explicit expressions of $E_{V,n}^{\text{NLO}}$ and $d_{V,n}^k$ can be found in the Appendix of Ref.~\cite{Wang:2017ijn}.

To construct the sum rules for the form factors $f_{\rm LP}^{+,-}(q^2)$, we express the correlation function in the form of a dispersion integral. Since the correlation function is given by the convolution of the hard kernels and the $D$-meson LCDAs, the imaginary parts of the hard kernels in Eqs. (\ref{hk1}) and  (\ref{hk2}) are essential for constructing the dispersion integral. Given that our hard kernels are closely analogous to those for the \(B\to\pi\) transition, the imaginary part of the correlation function can be extracted from Eqs. (B.2) and (B.20) in \cite{Duplancic:2008ix}. Substituting the derived imaginary parts of $\Pi(s)$ and $\tilde{\Pi}(s)$ into Eq. (\ref{sumrulemain}), we finally obtain the desired sum rules, which can be written as $f_{\rm LP}^{+,-,(1)}(q^2)$.

\subsection{The power-suppressed contributions}

In the $B$-meson decay processes, power-suppressed contributions generally play an essential role due to the finite mass of the b quark. In this work, we incorporate three categories of next-to-leading power(NLP) corrections: the charm-mass effect arising from the charm-quark propagator, the correction induced by the finite $D$-meson mass, and the two-particle twist-3 and twist-4 contributions.

\subsubsection{Charm-mass term in the bottom-quark propagator}
At NLO, diagrams (a) and (d) in Fig.~\ref{fig:NLO} involve the charm-quark propagator. The charm-quark mass term is suppressed by a factor of \(m_c/n\cdot p\) and is thus omitted in the leading-power approximation. For subleading-power contributions, however, the charm-mass effect must be incorporated. After performing the Lorentz structure reduction for diagrams (a) and (d) in Fig.~\ref{fig:NLO}, we derive the matrix element of the operator \(O_1(v)\), which reveals that these contributions also involve the twist-2 LCDA of the $D$ meson. Accordingly, the factorization formula for this term takes the following form
\begin{eqnarray}
	&&	\Pi_{\rm NLP, m_c}^{(1)} =\frac{\alpha_sC_F}{4\pi} f_D\bar n\cdot p m_c \int _0^1 du 	T^{(1)}_{1,m_c}(u,\mu)\phi_{D} (u, \mu)\, ,\nonumber \\ 
	&&	\tilde \Pi_{\rm NLP, m_c}^{(1)} =\frac{\alpha_sC_F}{4\pi} f_D\bar n\cdot p  m_c\int _0^1 du 	T^{(1)}_{2,m_c}(u,\mu)\phi_{D} (u, \mu)\, ,
	\label{NLOmc}
\end{eqnarray}
where the hard kernel
\begin{eqnarray}
	2T^{(1)}_{1,m_c}(u,\mu)&=&	T^{(1)}_{p,m_c}(u,\mu)+{\bar n \cdot q\over \bar n\cdot p}T^{(1)}_{q,m_c}(u,\mu) \nonumber \\ 
	2T^{(1)}_{2,m_c}(u,\mu)&=&	{ n \cdot q\over \bar n\cdot p}T^{(1)}_{q,m_c}(u,\mu) 
\end{eqnarray}

\begin{eqnarray}
	\begin{aligned}
		T^{(1)}_{p,m_c}(u,\mu)
		&= \frac{2}{\bar \rho m_b} \left( - \ln\frac{m_b^2}{\mu^2} + 1 \right)  + \frac{2}{u m_b (r_1 - r_2)} \frac{r_1}{\bar \rho} \Big[ {G}(r_1) - {G}(\rho) \Big] \\
		&\quad + \frac{2}{u m_b (r_1 - r_2)^2} \left[ \frac{\rho}{\bar u}\bar{G}(\rho) - \frac{u r_2}{\bar u}r{G}(r_2) - r_1 {G}(r_1) \right] \\
		&\quad + \frac{4}{u m_b (r_1 - r_2)^2} \left[ \frac{\rho}{\bar u}\ln\bar \rho - \frac{u r_2}{\bar u}\ln\bar r_2 - r_1 \ln\bar r_1 \right] \\
		&\quad + \frac{2}{u m_b (r_1 - r_2)\bar \rho} \left[ (2-r_1)\ln\bar r_1- \left( \frac{r_1}{\rho} + \bar \rho \right)\ln\bar \rho \right]. \\
			{T}^{(1)}_{q,m_c}(u,\mu)
		&= \frac{2 }{u m_b \bar \rho } \left[ G(r_1) - G(\rho) - \frac{\ln\bar \rho}{\rho} + \frac{\ln\bar r_1}{r_1} \right]
	\end{aligned}
\end{eqnarray}
As mentioned in the previous section, the sum rules for the contribution from charm mass effect  $f_{m_c}^{+,-,(1)}(q^2)$ can be obtained via Eq. (\ref{sumrulemain}) once the imaginary part of the hard kernels obtained, i.e.,
\begin{eqnarray}
	f^{+,(1)}_{m_c}(q^2)&=&\frac{f_D }{2\pi f_Bm_B^2}\int_0^1 du \int^{s_0^{\rm th}}_{m_b^2} ds{\rm exp}\left[-{s-m_B^2\over M^2}\right] \mathrm{Im}T^{(1)}_{p,m_c}(u,\mu)\phi_D(u,\mu), 
	\nonumber   \\ 
	f^{+,(1)}_{m_c}(q^2)+	f^{-,(1)}_{m_c}(q^2)&=&\frac{f_D }{\pi f_Bm_B^2}\int_0^1 du \int^{s_0^{\rm th}}_{m_b^2} ds{\rm exp}\left[-{s-m_B^2\over M^2}\right] \mathrm{Im}T^{(1)}_{q,m_c}(u,\mu)\phi_D(u,\mu),	
	\label{srmc}
\end{eqnarray}
where the imaginary part of the hard kernels is written as 
\begin{eqnarray}
	\begin{aligned}
	-\frac{1}{\pi} {\rm Im}T^{(1)}_{p,m_c}(u,\mu)
	&= \delta(-\bar \rho)\left\{
	-\frac{2}{m_b}\left(1-\ln\frac{m_b^2}{\mu^2}\right)
	\right. \\
	&\quad + \frac{2 r_1}{u m_b(r_1-r_2)}\left[
	\ln^2(-\bar r_2)-\frac{\pi^2}{6}-G(r_1)-\left(1-\ln\frac{m_b^2}{\mu^2}\right)\ln(-\bar r_2)
	\right] \\
	&\quad + \left.\frac{2}{u m_b(r_1-r_2)}\left[
	\left(\frac{r_1}{\rho}+\bar \rho\right)\ln(-\bar r_2)-(2-r_1)\ln(\bar r_1)
	\right]\right\} \\
	&\quad - \theta(-\bar \rho)\left\{
	\frac{2}{u m_b(r_1-r_2)}\left[\frac{r_1}{\rho}-\rho+\ln\frac{m_b^2}{\mu^2}+2\ln(-\bar \rho)\right]\frac{1}{-\bar \rho}\right. \\
	&\quad + \frac{2}{u m_b(r_1-r_2)^2}\frac{\rho}{\bar u}\left[
	2\ln(-\bar \rho)-\ln\rho+1+\ln\frac{m_b^2}{\mu^2}
	\right] \\
	&\quad - \frac{2 r_2}{\bar um_b(r_1-r_2)^2}\left[
	2\ln(-\bar r_2)-\ln r_2+1+\ln\frac{m_b^2}{\mu^2}
	\right]  \left.+ \frac{2 r_1}{u m_b(r_1-r_2)}\frac{\ln\rho}{\bar \rho}\right\} \\
	&\quad + \theta(\bar \rho)\frac{\alpha_s(\mu)C_F}{4\pi}\left\{
	-\frac{2 r_2}{\bar u m_b(r_1-r_2)^2}\left[
	2\ln(-\bar r_2)-\ln r_2+1+\ln\frac{m_b^2}{\mu^2}
	\right]
	\right\} \\ 
	-\frac{1}{\pi}\mathrm{Im}\tilde{T}^{(1)}_{q,m_c}(u)
	&= \frac{2\delta(\bar \rho)}{u m_b}\bigg[
	\ln^2(-\bar r_2)
	-\left(1-\ln\frac{m_b^2}{\mu^2}\right)\ln(-\bar r_2)
	-\frac{\pi^2}{6}
	-G(r_1)
	+\frac{\ln(-\bar r_2)}{\rho}
	-\frac{\ln\bar r_1}{r_1}
	\bigg] \\
	&\quad + \theta(-\bar \rho)\frac{2}{u m_b}
	\left[2\ln(-\bar \rho)-\ln\rho+\ln\frac{m_b^2}{\mu^2}-1+\frac{1}{\rho}\right]
	\left[\frac{1}{-\bar \rho}\right]_+.
	\end{aligned}
\end{eqnarray}

\subsubsection{Power corrections from $D$-meson mass}
Unlike the $B$-to-light-meson transition, the $D$-meson mass should not be neglected when we consider the power-suppressed contribution to $B \to D$ form factors. In the leading power contribution, the terms proportional to $p^2=m_D^2$ are dropped and they should be picked up at NLP. At leading order of the correlation function, this term  enters the denominator of the b-quark propagator. Explicitly, \(uP^2+\bar u q^2-m_b^2\) in the hard kernel becomes \(uP^2+ \bar u q^2- u\bar u m_D^2-m_b^2\) once the D-meson mass is taken into account, which modifies the denominator's pole structure and shifts the lower limit \(u_0\) in the sum rule \eqref{srlo}. The newly obtained $\tilde u_0$ reads
\begin{align}
	\tilde u_0 = \frac{1}{2 m_D^2} \left[ \sqrt{(s_0 - q^2 - m_D^2)^2 + 4 m_D^2(m_b^2 - q^2)} - (s_0 - q^2 - m_D^2) \right] \label{eq:u0}
\end{align}
Compared with the leading-power contribution in \eqref{srlo}, the power-suppressed contribution to the leading-order sum rules of $B\to D$ form factors induced by the $D$-meson mass can be written as
\begin{equation}
	f^{+,(0)}_{m_D}(q^2)=\frac{f_D m_b^2}{2f_Bm_B^2} \int _{\tilde u_0}^{u_0}  {du\over u} {\phi_{D} (u, \mu)} e^{(m_B^2-s(u))/M^2} . \label{srloNLP}
\end{equation}
We restrict our analysis to the leading-order contribution of this type of power correction, as higher-order corrections become quite complicated and are deferred to future work.

\subsubsection{Power corrections from two-particle twist-3 and twist-4 LCDAs}

Higher-twist LCDAs constitute an important class of contributions to heavy-to-light form factors. A systematic and consistent description of higher-twist corrections to the QCD correlation function requires accounting for the transverse-momentum dependence of valence (anti)quarks in the leading Fock-state wavefunctions, as well as contributions from higher Fock states. Nevertheless, higher-Fock-state LCDAs, which are generally numerically subdominant, are neglected in this work. Accordingly,  the two-particle LCDAs up to twist-4 are defined as follows.
\begin{align}
	\langle D(p) | \bar{c}^i_\alpha(x) q^j_\beta(0) | 0 \rangle
	&= \frac{i\delta^{ij}}{12} f_D \int_0^1 du \, e^{i u p\cdot x}
	\Bigg(
	[\slashed{p}\gamma_5]_{\beta\alpha} \phi_D(u)
	- [\gamma_5]_{\beta\alpha} \mu_D \phi_{3D}^p(u)
	+\nonumber\\
	&\quad \frac16 [\sigma_{\rho\tau}p^\rho x^\tau\gamma_5]_{\beta\alpha}  \mu_D \phi_{3D}^\sigma(u)
	+ \frac{1}{16} [\slashed{p}\gamma_5]_{\beta\alpha} x^2 \phi_{4D}(u)
	- \frac{i}{2}[\slashed{x}\gamma_5]_{\beta\alpha} \int_0^u \psi_{4D}(v) dv
	\Bigg),
\end{align}
where $\phi_{3D}^p(u)$ and $\phi_{3D}^\sigma(u)$ denote the twist-3 LCDAs of the $D$ meson, $\phi_{4D}(u)$ and $\psi_{4D}(u)$ stand for the twist-4 LCDAs, and $\mu_D=m_D^2/(m_c+m_q)$. Evaluating the tree-level Feynman diagrams in Fig.~\ref{fig:LO} straightforwardly
leads to the factorization formula of the higher-twist LCDA contribution to the correlation function,

\begin{equation}
	\begin{aligned}
		\Pi_{\rm NLP, ht}&= {1\over 2}\bar n\cdot p \left[A_0+A_1p\cdot q \right]+{1\over 2}\bar n\cdot q \left[B_0+B_1p\cdot q \right], \\
		\tilde \Pi_{\rm NLP, ht}&= {1\over 2} n\cdot p \left[A_0+A_1p\cdot q \right]+{1\over 2} n\cdot q \left[B_0+B_1p\cdot q \right],
		\end{aligned}	
\end{equation}
where 
\begin{align}
	A_0
	=&
	{f_D\mu_D\over  m_b} \int_0^1 du\,\left[\frac{6u\phi_{3D}^p(u)+\phi_{3D}^\sigma(u)}{6(\bar \rho+u\bar u r_D)}+\frac{(\bar u^2r_D)\phi_{3D}^\sigma(u)}{(\bar \rho+u\bar u r_D)^2}\right]\\
	-&\frac{f_D }{2m_b^2}\int_0^1 du\,\left[\frac{\phi_{4D}(u)}{(\bar \rho+u\bar u r_D)^3}-\frac{2}{(\bar \rho+u\bar u r_D)^2}\int_0^u \psi_{4D}(v)dv\right],\nonumber \\[4pt]
	A_1 	=&
	-{f_D\mu_D\over  3m_b^3}\int_0^1 du\,\frac{u\phi_{3D}^\sigma(u)}{(\bar \rho+u\bar u r_D)^2},\nonumber\\[4pt]
	B_0
	=&
{f_D\mu_D\over  m_b}  \int_0^1 du\,\left[\frac{u\phi_{3D}^p(u)}{(\bar \rho+u\bar u r_D)}-\frac{ur_D\phi_{3D}^\sigma(u)}{(\bar \rho+u\bar u r_D)^2}\right]-\frac{f_D }{m_b^2}\int_0^1 du\,\frac{1}{(\bar \rho+u\bar u r_D)^2}\int_0^u \psi_{4D}(v)dv,\nonumber\\[4pt]
	B_1
	=&
	-{f_D\mu_D\over  3m_b^3}\int_0^1 du\,\frac{\phi_{3D}^\sigma(u)}{(\bar \rho+u\bar u r_D)^2},
\end{align}
where $r_D=m_D/m_B$. Expressing the above result as a dispersion integral, we perform the Borel transformation to suppress the contributions from the higher excited and continuum states. By further invoking quark-hadron duality, we extract the desired form factor as 
\begin{eqnarray}
	f^{+,(0)}_{\rm ht}(q^2)&=&\frac{f_D m_b e^{m_B^2/M^2}}{2f_Bm_B^2} \bigg\{\int _{ u_0}^{1}  {du\over u}  e^{-s(u)/M^2}\left[\rho_1(u,q^2)+{1\over M^2}\rho_2(u,q^2)+{1\over 2M^4}\rho_3(u,q^2)\right] \nonumber \\ &&+e^{-s_0/M^2}\left[\rho_4(u_0,q^2)+{1\over M^2}\rho_5(u_0,q^2)\right]\bigg\}, \nonumber \\
	f^{+,(0)}_{\rm ht}(q^2)+	f^{-,(0)}_{\rm ht}(q^2)&=&\frac{f_D m_b e^{m_B^2/M^2}}{f_Bm_B^2} \bigg\{\int _{ u_0}^{1}  {du\over u}  e^{-s(u)/M^2}\left[\tilde\rho_1(u,q^2)+{1\over M^2}\tilde\rho_2(u,q^2)\right] \nonumber \\ &&+e^{-s_0/M^2}\left[\tilde\rho_4(u_0,q^2)\right]\bigg\}, \label{srht}
\end{eqnarray}
where 
\begin{eqnarray}
	\rho_1(u,q^2)&=&\mu_D\left[u\phi_{3D}^p(u)+{1\over 2}\phi_{3D}^\sigma(u)\right],\,
	\rho_2(u,q^2)={1\over 6}\mu_D{\bar u \phi_{3D}^\sigma(u) }\left[s(u)-m_D^2+\frac{2-u}{u}q^2\right]+{\bar um_b\over 2u}\phi_{4D}(u)\nonumber \\ 
	\rho_3(u,q^2)&=&{m_b\over2}{\bar u^2\over u^2}q^2\phi_{4D}(u),\, \rho_5(u_0,q^2)={m_b\over4}{\bar u_0^2 \phi_{4D}(u_0)\over u_0^2 D(u_0,q^2)}q^2,\nonumber \\
	\rho_4(u_0,q^2)&=&{\mu_D\over6}{\bar u_0 \phi_{3D}^\sigma(u_0)\over  D(u_0,q^2)}\left[s_0-m_D^2+\frac{2-u_0}{u_0}q^2\right]+{\bar u_0 m_b\over 2u_0}\left[{1\over D(u_0,q^2)}+{q^2\over 2D^2(u_0,q^2)}\right]\phi_{4D}(u_0),
	\end{eqnarray}
where $D(u,q^2)=(2u-1)m_D^2+s(u)-q^2$ , and 
\begin{eqnarray}
	\tilde\rho_1(u,q^2)&=&\mu_D\phi_{3D}^p(u),\,
	\tilde\rho_2(u,q^2)=-{1\over 6}\mu_D{\bar u \phi_{3D}^\sigma(u) }{D(u,q^2)\over u}-{\bar um_b\over u}\int_0^u\psi_{4D}(v)dv\nonumber \\ 
	\tilde\rho_4(u_0,q^2)&=&-{1\over6}\mu_D{\bar u_0 \phi_{3D}^\sigma(u_0)\over u_0 }-{\bar u_0m_b \over u_0 D(u_0,q^2)}\int_0^{u_0}\psi_{4D}(v)dv.
	\end{eqnarray}
Summing all the contributions considered above, we obtain our final analytical results for the LCSR predictions of the $B\to D$ form factors.
\begin{eqnarray}
f^+(q^2)&=&f_{\rm LP}^{+,(0)}(q^2)+f_{\rm LP}^{+,(1)}(q^2)+f_{\rm ht}^{+,(0)}(q^2)+f_{\rm m_D}^{+,(0)}(q^2)+f_{\rm m_c}^{+,(1)}(q^2),\nonumber \\
f^+(q^2)+f^-(q^2)&=&f_{\rm LP}^{+,(1)}(q^2)+f_{\rm LP}^{-,(1)}(q^2)+f_{\rm ht}^{+,(0)}(q^2)+f_{\rm ht}^{-,(0)}(q^2)+f_{\rm m_c}^{+,(1)}(q^2)+f_{\rm m_c}^{-,(1)}(q^2).
\end{eqnarray}
\section{Numerical analysis}

This section presents the numerical evaluation of the LCSR for the $B\to D$
form factors derived in Sec.~2. We specify the input parameters and the
$D$-meson light-cone distribution amplitudes (LCDAs) in
Sec.~\ref{sec:inputs}, discuss the resulting form factors and their
uncertainty budget in Sec.~\ref{sec:ff}, describe the BGL parametrisation and
the combined fit to the LCSR and lattice data in Sec.~\ref{sec:bgl}, and
finally extract $|V_{cb}|$ together with the lepton-flavour universality
ratio $R(D)$ in Sec.~\ref{sec:vcb}.

\subsection{Input parameters and the $D$-meson LCDAs}
\label{sec:inputs}

\begin{figure}[h]
	\begin{center}
		\vspace{0.3cm}\hskip0cm
		\includegraphics[width=0.5\textwidth]{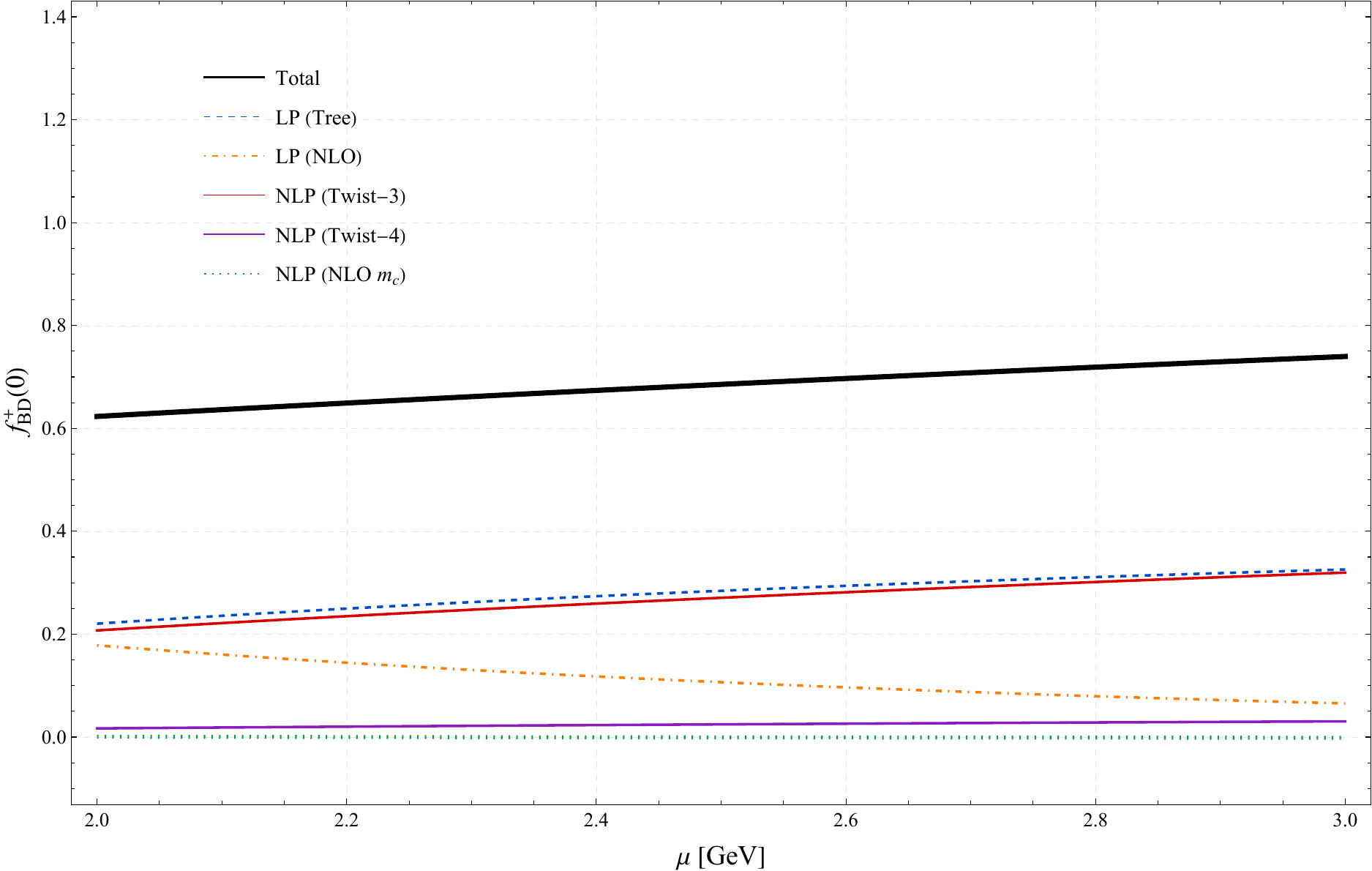}
	\end{center}
	\caption{ {Factorization scale $\mu$ dependence of the form factor $f_{BD}^{+}(q^2=0)$ in the leading-power approximation. A comment on the nomenclature is in order: ``LP (Tree)'', ``LP (NLO)'', ``NLP (Twist-3)'', ``NLP (Twist-4)'' and ``NLP (NLO $m_c$)'' correspond to the leading-power contribution at tree level, the leading-power contribution at next-to-leading order, the next-to-leading-power contributions from the twist-3 LCDAs, from the twist-4 LCDAs and from the charm-quark mass correction, respectively.}}
	\label{fig:scale}
\end{figure}

The QCD and hadronic parameters entering the sum rules are collected in
Table~\ref{tab:parameters}. The meson masses $m_B$ and $m_D$ are taken from
the experimental averages and enter as purely kinematic quantities, whereas
the $\overline{\rm MS}$ quark masses $m_b(m_b)$ and $m_c(m_c)$ control the
short-distance dynamics of the correlation function and, through the
prefactor $f_Bm_B^2/m_b$, the normalisation of the sum rule. The decay
constants $f_B$ and $f_D$ are taken from high-precision lattice
determinations and the latter fixes the normalisation of the $D$-meson LCDAs entering
through the $B$-meson interpolating current. The factorisation scale
$\mu=2.4$~GeV is chosen in the window between $\Lambda_{\rm QCD}$ and the
heavy-quark masses, where both the perturbative hard kernel and the evolution
of the LCDAs are under control and the residual scale dependence is
minimised once the next-to-leading-order coefficient function is included;
the size of the residual dependence is illustrated in Fig.~\ref{fig:scale}.
The LCDAs are evolved from the initial scale $\mu_0=m_D$, at
which the lattice determination of the $D$-meson distribution amplitude is
performed, up to $\mu$ by means of the ERBL evolution, so that the matching of
the quasi-distribution amplitude and the resummation of the endpoint
logarithms are carried out consistently at the scale at which the lattice
input is defined.

\begin{figure}[h]
	\begin{center}
		\vspace{0.3cm}\hskip0cm
		\includegraphics[width=0.5\textwidth]{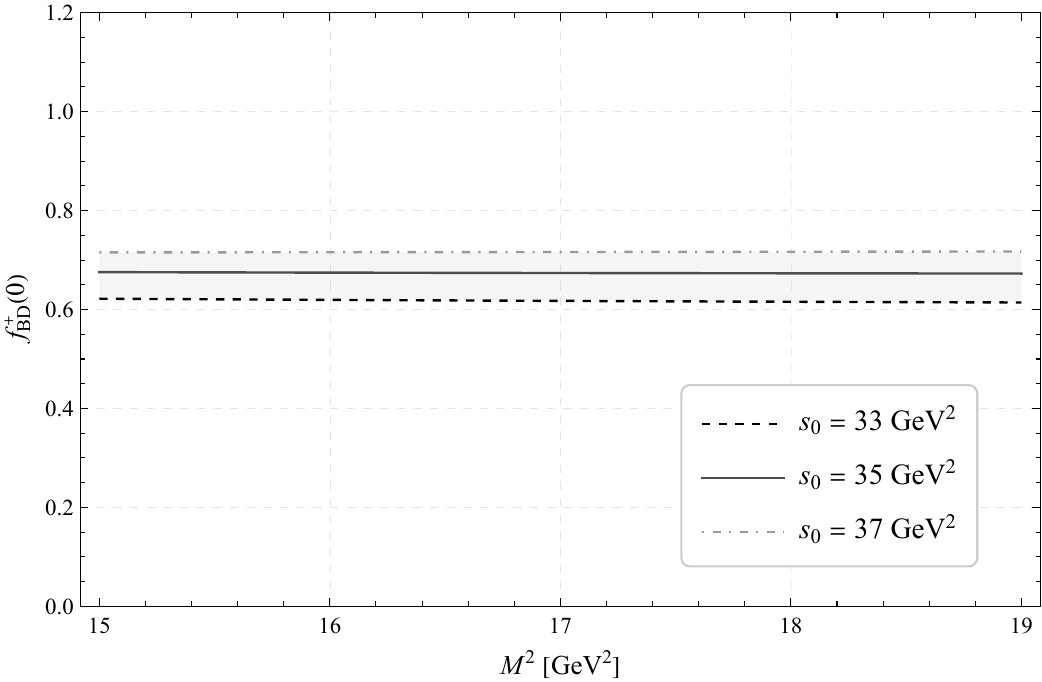}
	\end{center}
	\caption{ Dependence of the form factors on the Borel mass. }
	\label{fig:Borel}
\end{figure}

\begin{figure}[h]
\begin{center}
	\vspace{0.3cm}\hskip0cm
	\includegraphics[width=0.5\textwidth]{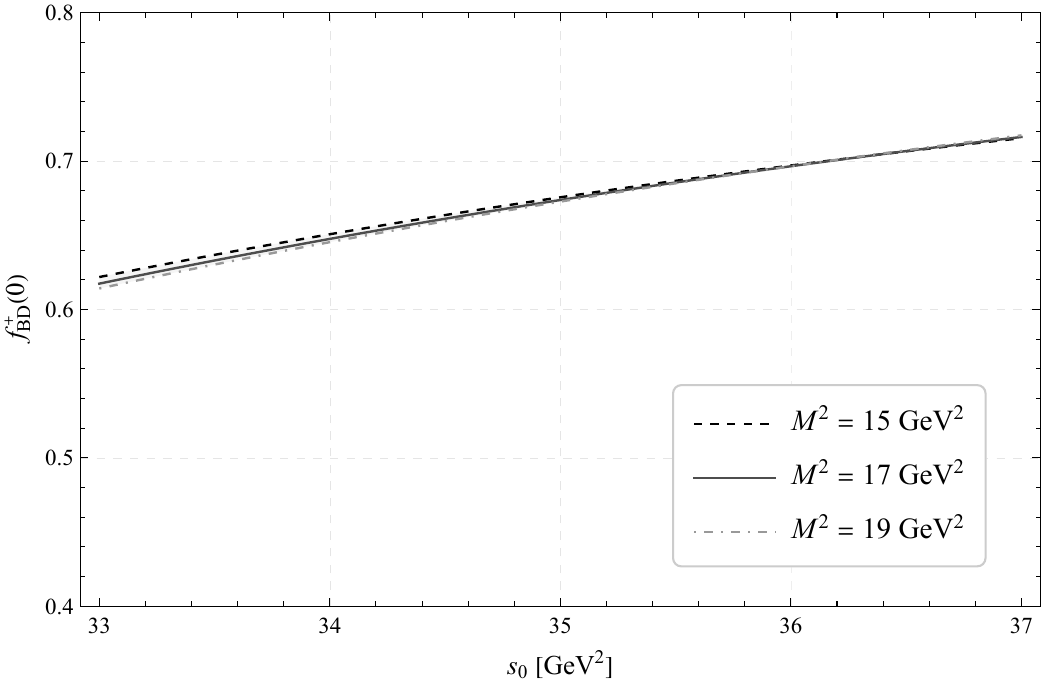}
\end{center}
\caption{ Dependence of the form factors on the threshold parameter. }
\label{fig:threshold}
\end{figure}

The Borel mass $M^2=17\pm2.0$~GeV$^2$ and the continuum threshold
$s_0=35(2)$~GeV$^2$ lie inside the conventional working window of the sum
rules. The lower limit on $M^2$ is fixed by the requirement that the
higher-resonance and continuum contributions, estimated in the
parton-hadron-duality approximation, stay below the level of the power
corrections retained in the OPE; the upper limit is fixed by the requirement
that the truncated OPE still converges. Within this window the form factors
depend only mildly on $M^2$ and on $s_0$, as is explicitly shown in
Figs.~\ref{fig:Borel} and~\ref{fig:threshold}, where the variation of
$f_+^{BD}(0)$ with the Borel parameter, and with the threshold, is displayed
for three values of the other parameter. The residual sensitivity found
there, of the order of $2$--$3\%$, is included in the error budget discussed
below.

The non-perturbative input specific to the present calculation is the
$D$-meson LCDA set. For the leading-twist distribution amplitude
$\phi(u,\mu)$ we use the first-principles determination obtained from lattice
QCD within the HQLaMET framework. In this approach the hierarchy
$\Lambda_{\rm QCD}\ll m_H\ll P^z$ allows one to construct a
lattice-computable quasi-DA from equal-time correlators of a boosted $D$
meson, which is then matched to the LCDA by a perturbative kernel. After
hybrid renormalisation to remove the ultraviolet divergences associated with
the Wilson lines, a $\lambda=zP_z$ extrapolation is applied to reconstruct the
long-distance tail of the coordinate-space matrix elements; subsequent
Fourier transformation, LaMET matching with endpoint-logarithm resummation and
ERBL evolution are carried out, followed by continuum, chiral and
infinite-momentum extrapolations to obtain the physical QCD LCDA at
$\mu=m_D$. The resulting $\phi(u,\mu=m_D)$ peaks at $u\approx0.2$--$0.3$ and
falls slowly towards $u\to1$, with total uncertainties below $30\%$ in the
region $0.1<u<0.9$; the endpoints are not first-principles predictions. The
Gegenbauer moments are stable against the truncation order, and the optimal
fitted values read $a_1^D=0.449(38)$ and $a_2^D=0.146(33)$, in agreement
with the lattice-OPE results $a_1^D=0.434(17)$ and $a_2^D=0.183(73)$. Since
the OPE route requires no large boost, this agreement demonstrates that the
finite-$P^z$ systematics of the LaMET reconstruction are under control.

For the twist-3 and twist-4 LCDAs, there exist very few studies in the literature. Fortunately, they are not independent and are related by the QCD equations of motion. The relations for the non-local operators read \cite{Ball:1998sk,Ball:1999aj}
\begin{align}
	\frac{\partial}{\partial x_\mu} \bar{c}(x)\gamma_\mu\gamma_5 q(0)
	&= -i \int_{0}^{1} dv\, v \bar{c}(x) x_\alpha g G_{\alpha\mu}(vx) \gamma_\mu\gamma_5 q(0) + m_c\bar{q}(x)i\gamma_5 q(0), \nonumber \\[4pt]
	\partial_\mu\bigl\{\bar{q}(x)\gamma_\mu\gamma_5 q(0)\bigr\}
	&= -i \int_{0}^{1} dv\, \bar{c}(x) x_\alpha g G_{\alpha\mu}(vx) \gamma_\mu\gamma_5 q(0) +m_c\bar{c}(x)i\gamma_5 q(0), \nonumber\\[4pt]
	\partial_\mu\bar{c}(x)\sigma_{\mu\nu}\gamma_5q(0)
	&= -i \frac{\partial}{\partial x_\nu} \bar{c}(x)\gamma_5 q(0)
	+ \int_{0}^{1} dv\, v\bar{q}(x) x_\rho g G_{\rho\nu}(vx)\gamma_5 q(0) \nonumber\\
	&\quad -i \int_{0}^{1} dv\, \bar{c}(x) x_\rho g G_{\rho\mu}(vx)\sigma_{\mu\nu}\gamma_5 q(0) -m_c\bar{c}(x)\gamma_\nu\gamma_5 q(0),
	\nonumber\\
	\frac{\partial}{\partial x_\mu} \bar{c}(x)\sigma_{\mu\nu}\gamma_5 q(0)
	&= -i\partial_\nu \bar{c}(x)\gamma_5 q(0)
	+ \int_{0}^{1} dv\, \bar{c}(x) x_\rho g G_{\rho\nu}(vx)\gamma_5 q(0) \nonumber\\
	&\quad -i \int_{0}^{1} dv\, v\bar{c}(x) x_\rho g G_{\rho\mu}(vx)\sigma_{\mu\nu}\gamma_5 q(0)- m_c\bar{q}(x)\gamma_\nu\gamma_5 q(0).
	\label{eq:eom}
\end{align}
Note that 
\begin{align}
	\frac{\partial}{\partial x_\mu} \bar{c}(x)\Gamma  q(0)
	&= \left[\frac{\partial}{\partial x_\mu} \bar{c}(x)\right]\Gamma  q(0)-\bar{c}(x)\Gamma  \left[\frac{\partial}{\partial x_\mu} q(x)\right]_{x=0}
\end{align}
and 
\begin{align}
\partial_\mu(x)[\bar{c}(x)\Gamma  q(0)]={\partial \over \partial y^\mu}[\bar{c}(x+y)W(x+y,y)\Gamma  q(y)]_{y=0}
\end{align}
where $W(x,y)$ denotes the Wilson lines that keep the non-local operators gauge invariant. Neglecting the contribution from three-particle LCDAs, we can express the twist-3 LCDAs in terms of the twist-2 ones \cite{Ball:2006wn}.
\begin{eqnarray}
	\phi_{3}^p(u)
	&\simeq& 1 + 3\rho_+^D(1 + 6a_2^D) - 9\rho_-^D a_1^D
	+ C_1^{1/2}(2u-1)
	\left[
	\frac{27}{2}\rho_+^D a_1^D - \rho_-^D\left(\frac{3}{2}+27a_2^D\right)
	\right] \nonumber\\
	&+& C_2^{1/2}(2u-1)\left( 15\rho_+^D a_2^D - 3\rho_-^D a_1^D\right)
	+ C_3^{1/2}(2u-1)\left(- \frac{9}{2}\rho_-^D a_2^D\right) \nonumber\\
	&+& 
	\frac{3}{2}\left(\rho_+^D+\rho_-^D\right)\left(1-3a_1^D+6a_2^D\right)\ln u
	+\frac{3}{2}\big(\rho_+^D-\rho_-^D\big)\big(1+3a_1^D+6a_2^D\big)\ln\bar{u},\\
	\phi_{3}^\sigma(u)
	&\simeq&6u\bar{u}\bigg[
	1+\frac{3}{2}\rho_+^D+15\rho_+^D a_2^D-\frac{15}{2}\rho_-^D a_1^D
	+\left(3\rho_+^D a_1^D-\frac{15}{2}\rho_-^D a_2^D\right)C_1^{3/2}(2u-1) 
	\nonumber\\
	&+&
	\left(\frac32\rho_+^D a_2^D\right)C_2^{3/2}(2u-1)
	\bigg]  +9u\bar{u}(\rho_+^D+\rho_-^D)(1-3a_1^D+6a_2^D)\ln u
	\nonumber\\	&+&9u\bar{u}(\rho_+^D-\rho_-^D)(1+3a_1^D+6a_2^D)\ln\bar{u}, 
\end{eqnarray}
where \begin{align}
\rho_+^D = \frac{(m_c + m_q)^2}{m_D^2},\qquad \rho_-^D = \frac{m_c^2 - m_q^2}{m_D^2}. 
\end{align}
We neglect the light-quark masses, as they are much smaller than the charm-quark mass. 
	The twist-4 LCDAs can be separated into ``genuine'' twist-4 contributions $\psi_{4;D}^{T4},\phi_{4;D}^{T4}$ and Wandzura-Wilczek-type mass corrections $\psi_{4;D}^{WW},\phi_{4;D}^{WW}$ \cite{Ball:2006wn}, i.e., 
\begin{align}
	\psi_{4;D}(u) = \psi_{4;D}^{T4}(u) + \psi_{4;D}^{WW}(u),\quad \phi_{4;D}(u) = \phi_{4;D}^{T4}(u) + \phi_{4;D}^{WW}(u).
\end{align}
with
\begin{align}
	\psi_{4;D}^{T4}(u) &= \frac{20}{3}\delta_D^2 C_2^{1/2}(2u-1) + 5\left\{5\theta_1^D-\theta_2^D\right\}C_3^{1/2}(2u-1), \label{eq:4.30} \\
	\psi_{4;D}^{WW}(u)
	&\simeq m_D^2\left\{1 + 6\rho_+^D(1+6a_2^D)-18\rho_-^D a_1^D\right\}C_0^{1/2}(2u-1) \notag \\
	&\quad + m_D^2\left\{-12\kappa_{4D}-\frac95 a_1^D +27\rho_+^D a_1^D -3\rho_-^D(1+18a_2^D)\right\}C_1^{1/2}(2u-1) \notag \\
	&\quad + \left\{m_D^2\left(1+\frac{18}{7}a_2^D+30\rho_+^D a_2^D -6\rho_-^D a_1^D\right)\right\}C_2^{1/2}(2u-1) \notag \\
	&\quad+ \left\{ m_D^2 \left(\frac{9}{5} a_1^D + \frac{16}{3}\kappa_{4D} - 9\rho_{-}^D a_2^D\right) \right\} C_3^{1/2}(2u - 1) \notag \\
	&\quad + \left\{-\frac{9}{28}m_D^2 a_2^D \right\}C_4^{1/2}(2u-1) 
	+ 6m_c^2(1-3a_1^D+6a_2^D)\ln u,
\end{align}
and 
\begin{align}
	\phi_{4;D}^{T4}(u)
	&\simeq \frac{200}{3}\delta_D^2 u^2\bar{u}^2 + 20u^2\bar{u}^2\left\{4\theta_1^D - 5\theta_2^D\right\} \notag \\
	&\quad + 40\phi_2^D\left\{u\bar{u}(2u-1)(2-3u\bar{u})
	-\left[2u^3(u-2)\ln u\right] + [u\leftrightarrow\bar{u}]\right\},
\\
	\phi_{4;D}^{WW}(u)
	&\simeq \frac{16}{3}m_D^2\kappa_{4D}\left\{u\bar{u}(2u-1)(1-2u\bar{u})
	+\left[5(u-2)u^3\ln u\right] - [u\leftrightarrow\bar{u}]\right\} \notag \\
	&\quad -\frac{36}{5}m_D^2 a_2^D
	\left\{
	\frac14 u\bar{u}(4-9u\bar{u}+110u^2\bar{u}^2)
	+\left[u^3(10-15u+6u^2)\ln u\right] + [u\leftrightarrow\bar{u}]
	\right\} \notag \\
	&\quad + 4m_D^2 u\bar{u}(1+3u\bar{u})\left[(1+\frac95 a_1^D (2u-1)\right]. \label{eq:4.34}
\end{align}
where the parameters $\delta_D^2$ and $\kappa_{4D}$ are defined as 
\begin{align}
	\langle   D(P)| \bar{c} \gamma_\alpha g \widetilde{G}_{\mu\alpha} q |0 \rangle
	&= -i P_\mu f_D \delta_D^2, \notag\\
	\langle  D(P)| \bar{c} \gamma_\alpha \gamma_5 i g G_{\mu\alpha} q |0  \rangle
	&= -i P_\mu f_D m_D^2 \kappa_{4D}.
\end{align}
In the above equations, we have omitted the parameters entering the twist-3 three-particle distribution amplitudes. $\delta_D^2$ is then evaluated from QCD sum rules, whose analytic expression reads \cite{Novikov:1984af,Ball:2006wn}.
\begin{align}
	f_D^2 \delta_D^4 e^{-m_D^2/M^2}
	&= \frac{\alpha_s}{160\pi^3}\,(M^2)^3\left(\ln\frac{\mu^2}{M^2}-\gamma_E\right) + \frac{1}{72}\left\langle\frac{\alpha_s}{\pi}G^2\right\rangle\left(\ln\frac{\mu^2}{M^2}-\gamma_E\right) \notag\\
	&\quad - \frac{2\alpha_s m_c \langle\bar q q\rangle}{9\pi}\left(\ln\frac{\mu^2}{M^2}-\gamma_E\right) + \frac{\alpha_s m_c \langle\bar q\sigma g G q\rangle}{9\pi M^2}\left(\ln\frac{\mu^2}{M^2}-\gamma_E -1\right) \notag\\
	&\quad - \frac{2\alpha_s m_c \langle\bar q\sigma g G q\rangle}{27\pi M^2}  - \frac{25\pi\alpha_s\,m_0^2 \langle\bar q q\rangle^2}{324\,(M^2)^2}.
\end{align}
The numerical calculation results in $\delta_D^2=0.107\pm0.021$.  The value of $\kappa_{4D}$ can be obtained by equations of motion, which results in 
\begin{align}
	\kappa_{4D} \equiv -\frac{1}{8}\, - \frac{9}{40}\,a_1^D + \frac{m_c^2 }{2 m_D^2}.
\end{align}
In addition 
\begin{align}
	 \theta_1^D &= \frac{7}{10}a_1^D \delta_D^2, & \theta_2^D &= -\frac{7}{5}a_1^D \delta_D^2, & \phi_2^D &= -\frac{7}{20}a_1^D \delta_D^2. 
	\end{align}

\begin{table*}[htbp]
	\centering
	\renewcommand{\arraystretch}{1.2} 
	\setlength{\tabcolsep}{6pt}      
	\begin{tabular}{|l l l | l l l|}
	\hline
	Parameters & values & ref. & Parameters & values & ref. \\
	\hline
	$m_{B}$ & 5.279 GeV & \cite{ParticleDataGroup:2024cfk} & $m_{D}$ & 1.869 GeV & \cite{ParticleDataGroup:2024cfk} \\
	\hline
	$m_b(m_b)$ & $4.196\pm0.009$ GeV & \cite{ParticleDataGroup:2024cfk} & $m_c(m_c)$ & $1.28\pm0.025$ GeV & \cite{ParticleDataGroup:2024cfk} \\
	\hline
	$f_B$ & $(190.0 \pm 1.3)$ MeV & \cite{FlavourLatticeAveragingGroupFLAG:2021npn} & $f_D$ & $212.0(7)$ MeV & \cite{FlavourLatticeAveragingGroupFLAG:2021npn} \\
	\hline
	$\mu$ & $2.4$ GeV &  & $\mu_0$ & $1.869$ GeV &  \\
	\hline
	$M^2$ & $17 \pm 2.0$ GeV$^2$ &  & $s_0$ & $35\pm 2.0$ GeV$^2$ &  \\
	\hline
	
	\end{tabular}
	\caption{Input parameters used in the numerical calculation of the form factors.}
	\label{tab:parameters}	
\end{table*}

\begin{table}[h]
	\centering
	
	\renewcommand{\arraystretch}{1.8} 
	\setlength{\tabcolsep}{8pt}
	
	\begin{tabular}{|c|cccccc|}
		\hline
		$q^2 \ [\text{GeV}^2]$ & $-3.0$ & $-2.0$ & $-1.0$ & $0.0$ & $1.0$ & $2.0$ \\ \hline
		$f_{BD}^+$ & $0.597^{+0.049}_{-0.060}$ & $0.621^{+0.050}_{-0.061}$ & $0.646^{+0.050}_{-0.062}$ & $0.674^{+0.051}_{-0.063}$ & $0.703^{+0.052}_{-0.064}$ & $0.735^{+0.052}_{-0.065}$ \\
		$f_{BD}^0$ & $0.626^{+0.052}_{-0.064}$ & $0.641^{+0.052}_{-0.063}$ & $0.657^{+0.051}_{-0.063}$ & $0.674^{+0.051}_{-0.063}$ & $0.691^{+0.051}_{-0.063}$ & $0.708^{+0.050}_{-0.062}$ \\ 
		\hline
	\end{tabular}
		\caption{Predictions for the form factors in the large-recoil region from LCSR.}
	\label{tab:LCSR-prediction}
\end{table}

 \begin{figure}[h]
	\begin{center}
		\vspace{0.3cm}\hskip0cm
		\includegraphics[width=0.5\textwidth]{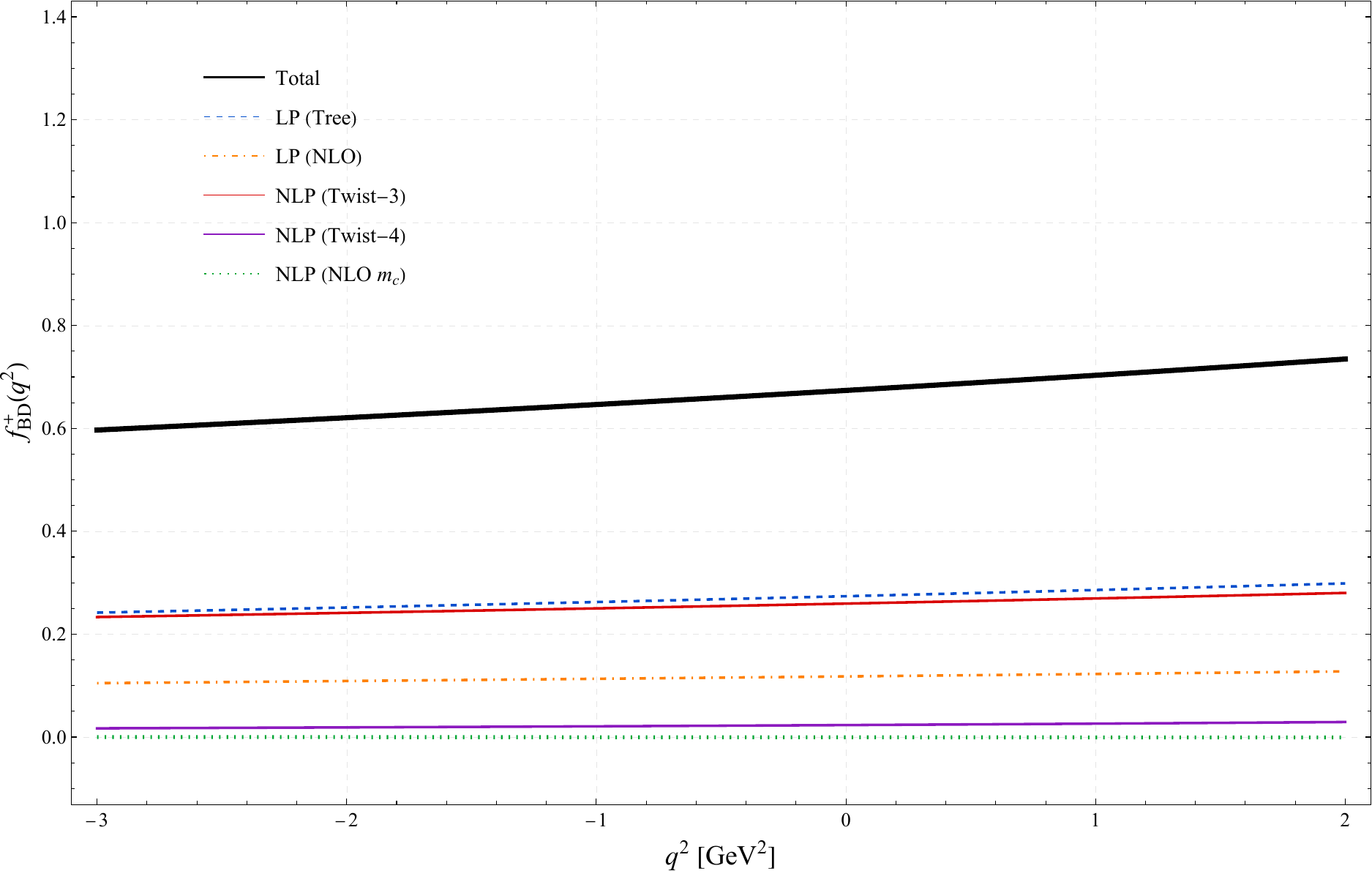}
	\end{center}
\caption{ {$q^2$ dependence of the form factor $f_{BD}^{+}(q^2)$ in the leading-power approximation. The NLP contribution from the higher-twist LCDAs is comparable to the leading-power contribution, while the charm-quark mass correction is negligible.}}
	\label{fig:q2dependence}
\end{figure}

\subsection{Form factors at large recoil}
\label{sec:ff}

The resulting form factors are displayed in Table~\ref{tab:LCSR-prediction}
for $-3~\mathrm{GeV}^2\le q^2\le 2~\mathrm{GeV}^2$. Both $f_+^{BD}(q^2)$ and $f_0^{BD}(q^2)$ rise
smoothly with $q^2$; the slope of the vector form factor is larger,
about $20\%$ over the range shown, while the scalar form factor is nearly
flat, rising by less than $10\%$. The two form factors coincide at $q^2=0$,
\begin{equation}
 f_+^{BD}(0)=f_0^{BD}(0)=0.674^{+0.051}_{-0.063} \,,
\end{equation}
as required by the kinematic identity $F^+(0)=F^0(0)$ following from the
Lorentz decomposition of the hadronic matrix element. This identity is satisfied automatically within our sum-rule framework. The quoted uncertainties, ranging from \(7\%\) to \(9\%\) and decreasing slowly with increasing \(q^2\), arise from several sources: the input uncertainties in the Gegenbauer moments \(a_1^D\) and \(a_2^D\), which enter the sum rules both through \(\phi(u)\) and through the equations-of-motion relations for the twist-three and twist-four distribution amplitudes; the residual dependence on the Borel mass and continuum threshold, as displayed in Figs.~\ref{fig:Borel} and \ref{fig:threshold}; the b- and c-quark masses; and the renormalisation-scale ambiguity, which is illustrated by the residual scale dependence in Fig.~\ref{fig:scale}, where the individual leading-power and NLP contributions are also shown separately. Compared with the B-meson light-cone sum-rule results of Ref.~\cite{Gao:2021sav}, our predictions have considerably smaller uncertainties because we employ LCDAs extracted from lattice-QCD simulations. It should be noted, however, that the present uncertainties are likely underestimated, since the higher-twist LCDAs used in this work are not yet complete. Although the central values show a sizable discrepancy, the two sets of results are consistent once the respective uncertainties are taken into account.

The decomposition of \(f_+^{BD}(q^2)\) into the individual terms of the factorization formula is shown in Fig.~\ref{fig:q2dependence}. In addition, we explicitly present the various contributions to \(f_+^{BD}(0)\) in Table~\ref{tab:contribution}. The leading-power tree-level amplitude provides the dominant contribution. The leading-power NLO correction is approximately \(18\%\), consistent with the expected \({\cal O}(\alpha_s/\pi)\) magnitude, and the sum of all leading-power contributions reaches nearly \(60\%\), dominating the form factor \(f_+^{BD}(0)\). The NLP contributions induced by twist-three distribution amplitudes are numerically comparable to their leading-power counterparts. This feature is peculiar to the heavy-to-heavy transition, for which \(m_D^2/m_B^2\) is not small and \(m_D/m_B\simeq0.35\). By contrast, the \({\cal O}(m_c/m_b)\) correction originating from the charm-mass term of the hard-collinear propagator turns out to be negligible within the current precision.

\begin{table}[htbp]
	\centering
	\caption{The contributions of different terms to $f_+^{B\to D}(0)$.}
	\label{tab:contribution}
	\begin{tabular}{c|c}
		\hline
			\textbf{contributions} & \textbf{central} \\
		\hline
		total & 0.6738 \\
		\hline
		twist-2 LO & 0.2737(40.62\%) \\
		\hline
		twist-2 NLO & 0.1178(17.48\%) \\
		\hline
		twist-3 LO & 0.2594(38.50\%) \\
		\hline
		$m_c$-NLO & -0.0004(-0.06\%) \\
		\hline
		twist-4 LO & 0.0233(3.46\%) \\
		\hline
	\end{tabular}
\end{table}

\subsection{BGL parametrisation and combined fit}
\label{sec:bgl}

The LCSRs determine the form factors only at large recoil,
$q^2\lesssim2$~GeV$^2$, whereas the lattice calculations cover the
small-recoil region $q^2\gtrsim8.5$~GeV$^2$; see Table~\ref{tab:lattice-data}.
To obtain the form factors over the whole physical range, and to combine the two sets of information
in a largely model-independent way, we use the BGL
parametrisation, which follows from the analyticity and unitarity properties
of the form factors and does not rely on any specific ansatz for their shape.
The BGL parametrisation introduces a function $z(q^2, t_0)$ that depends on the squared momentum transfer $q^2$ 
\begin{equation}
	z(q^2, t_0) = \frac{\sqrt{t_+ - q^2} - \sqrt{t_+ - t_0}}{\sqrt{t_+ - q^2} + \sqrt{t_+ - t_0}}, \qquad t_\pm = (m_B \pm m_D)^2,
\end{equation}
where $t_0$ is a free parameter. The threshold $t_+$ corresponds to the pair-production threshold, whereas $t_-$ is the physical kinematic endpoint in $b \to c$ semileptonic decays.
Introducing the Blaschke factor $P_i(z)$ and the outer function $\phi_i(z)$, the form factors can be parametrized as
\begin{equation}
	f_i(z) = \frac{1}{P_i(z)\,\phi_i(z)} \sum_{j=0}^{\infty} b_j^{(i)} z^j.
\end{equation}
The Blaschke factor $P_i$ removes the poles of the $B_c$ resonances lying below the pair-production threshold in the $q^2$ plane. In the $q^2$ variable it is defined as $P_i(q^2) = \prod_p z(q^2, m_p^2)$, while in the $z$ variable it takes the form
\begin{equation}
	P_i(z) = \prod_p \frac{z - z_p}{1 - z\, z_p}, \qquad z_p \equiv z(m_p^2, t_0),
\end{equation}
where the index $p$ runs over all resonances that share the same quantum numbers as the form factor $f_i$. The corresponding $B_c$ resonance masses $m_p$ are listed in Table \ref{tab:resonances}.
The outer functions   read
\begin{align}
	\phi_{f_+}(z) &= \frac{16 r^2}{m_B} \sqrt{\frac{2 n_I}{3\pi\,\tilde{\chi}_{1^-}^{T}(0)}}\,
	\frac{(1+z)^2(1-z)^{1/2}}{\left[(1+r)(1-z) + 2\sqrt{r}\,(1+z)\right]^5}, \\[6pt]
	\phi_{f_0}(z) &= 2r(1-r^2) \sqrt{\frac{2 n_I}{\pi\,\chi_{1^-}^{L}(0)}}\,
	\frac{(1+z)(1-z)^{3/2}}{\left[(1+r)(1-z) + 2\sqrt{r}\,(1+z)\right]^4},
\end{align} 
where the input parameters are the same as those in Ref.~\cite{Gao:2024vql}.  The coefficients are constrained by the
unitarity condition
\begin{equation}
\sum_{n=0}^{N}a_n^2\le1\,,\qquad\sum_{n=0}^{N}b_n^2\le1\,,
\end{equation}
which is far from being saturated in our fit, so that the truncated series
provides a reliable extrapolation to $q^2=0$ and to the kinematic endpoint.

\begin{figure}[h]
	\begin{center}
		\vspace{0.3cm}\hskip0cm
		\includegraphics[width=0.5\textwidth]{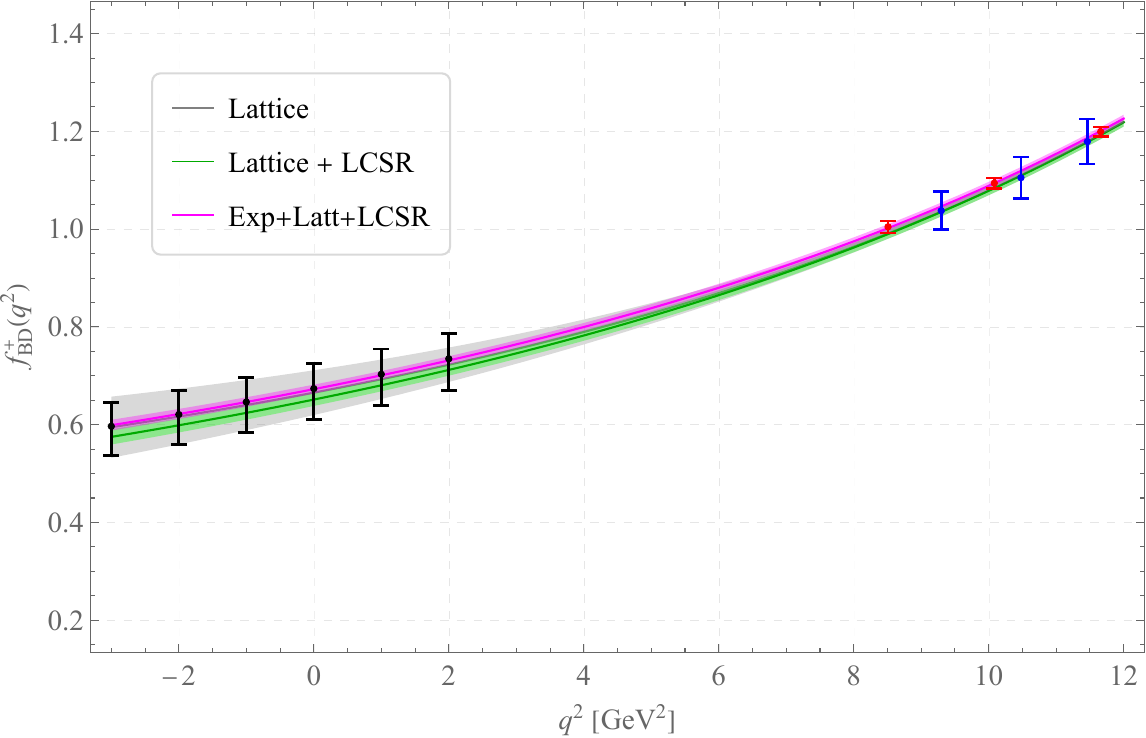}
	\end{center}
	\caption{BGL fitted results of the form factors as functions of $q^2$ . The red (blue) points with error bars represent the FNAL/MILC (HPQCD) data whereas the magenta points are LCSR results from this work for $f_{BD}^{+}(q^2)$. The magenta (green) curves correspond to $f_{BD}^{+}(q^2)$ obtained by fitting the combined data of LCSR and lattice predictions. The shaded regions are error bands due to uncertainties of the five parameters in Table~\ref{tab:correlation}. }
		\label{fig:fplus}
	\end{figure}

\begin{figure}[h]
		\begin{center}
			\vspace{0.3cm}\hskip0cm
			\includegraphics[width=0.5\textwidth]{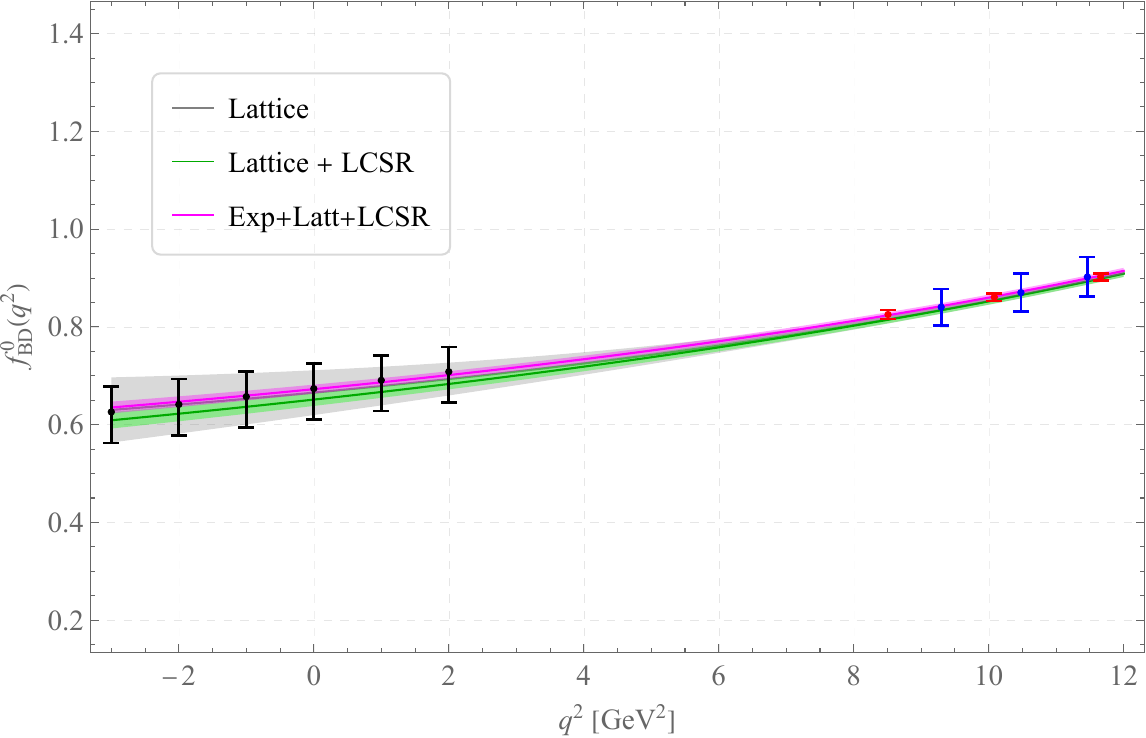}
		\end{center}
		\caption{BGL fitted results of the form factors as functions of $q^2$ . The red (blue) points with error bars represent the FNAL/MILC (HPQCD) data whereas the magenta points are LCSR results from this work for $f_{BD}^{0}(q^2)$. The magenta (green) curves correspond to $f_{BD}^{0}(q^2)$  obtained by fitting the combined data of LCSR and lattice predictions. The shaded regions are error bands due to uncertainties of the five parameters in Table~\ref{tab:correlation} }
			\label{fig:fzero}
		\end{figure}

We fit the five parameters $\{b_0^{f_+} , b_1^{f_+} , b_2^{f_+} , b_1^{f_0} , b_2^{f_0}\}$ to the six LCSR points of
Table~\ref{tab:LCSR-prediction} together with the FNAL/MILC and HPQCD data of
Table~\ref{tab:lattice-data}, by minimising
\begin{equation}
\chi^2=\sum_{i,j}\big(f_i^{\rm th}-f_i^{\rm data}\big)
\,C^{-1}_{ij}\big(f_j^{\rm th}-f_j^{\rm data}\big),
\end{equation}
where $C_{ij}$ denotes the covariance matrix, which includes both the
theoretical uncertainties of the LCSR predictions and the correlations among
the lattice data points of a given collaboration. The kinematic constraint
$f_+^{BD}(0)=f_0^{BD}(0)$, which must hold for any consistent set of form
factors, is imposed by eliminating $b_0^{f_0}$, so that the fit has five free
parameters. The resulting coefficients and their correlation matrix are given
in Table~\ref{tab:correlation}, and the corresponding shapes are shown in
Figs.~\ref{fig:fplus} and~\ref{fig:fzero} together with the input data. The
fit reproduces all data points within about $1\sigma$, and the unitarity sums
remain well below unity, which confirms that the truncation at $N=2$ is
adequate in the region $|z|\lesssim0.4$ probed by the data. 

The behaviour of the fitted form factors is physically sensible. The vector
form factor rises from $f_+^{BD}(0)=0.67$ to about $1.2$ at the kinematic
endpoint, with the bulk of the rise taking place for $q^2\gtrsim6$~GeV$^2$,
where the lattice data dominate the fit. The scalar form factor, in contrast,
stays close to its value at $q^2=0$ over the whole range, reflecting the
approximate heavy-quark symmetry relation between the two form factors. The
two data sets are mutually consistent: the LCSR points lie on the
extrapolation of the lattice-driven fit within errors, and the fit
interpolates smoothly between the region where the uncertainty is dominated by
the LCDA moments and the region where it is dominated by the lattice
statistics.

\subsection{Extraction of $|V_{cb}|$ and $R(D)$}
\label{sec:vcb}

\begin{figure}[h]
	\begin{center}
		\vspace{0.3cm}\hskip0cm
		\includegraphics[width=0.5\textwidth]{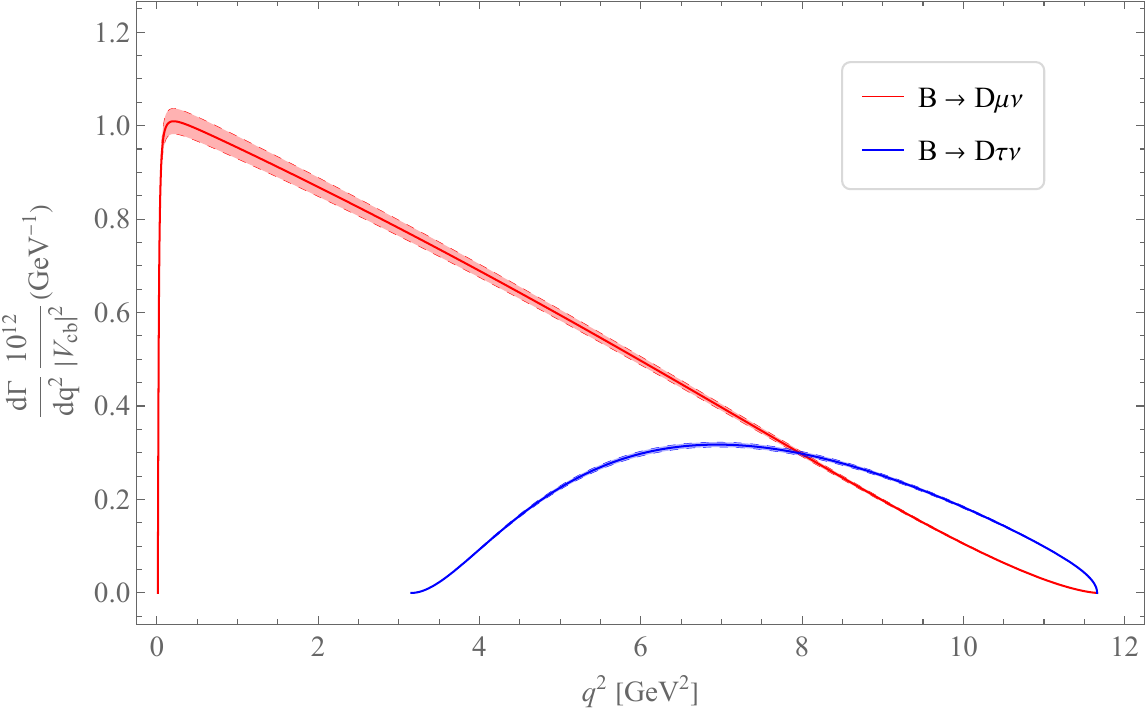}
	\end{center}
	\caption{ The differential decay rates for $B\to D\mu\nu_\mu$ (red curve) and $B\to D\tau\nu_\tau$ (blue curve) with error bands obtained from this work. The black points with error bars are the experimental data for $B\to D\mu\nu_\mu$ from the Belle collaboration. }
	\label{fig:BR}
\end{figure}

With the form factors at hand, the differential rate of the semileptonic decay
reads
\begin{equation}
\frac{d\Gamma}{dq^2}(B\to D\ell\nu)
=\frac{G_F^2|V_{cb}|^2}{192\pi^3m_B^3}\lambda^{3/2}(q^2)
\Big(1-\frac{m_\ell^2}{q^2}\Big)^2
\Big[\Big(1+\frac{m_\ell^2}{2q^2}\Big)\big|f_+^{BD}(q^2)\big|^2
+\frac{3m_\ell^2}{8q^2}\frac{(m_B^2-m_D^2)^2}{\lambda(q^2)}
\big|f_0^{BD}(q^2)\big|^2\Big],
\label{eq:rate}
\end{equation}
with $\lambda(q^2)=\big[q^2-(m_B-m_D)^2\big]\big[q^2-(m_B+m_D)^2\big]$. For
$\ell=\mu$ the second term is strongly suppressed by $m_\mu^2/q^2$ and the
spectrum is governed by $|f_+^{BD}(q^2)|^2$ alone, whereas for $\ell=\tau$
both form factors contribute with comparable weight and the phase space is
reduced by more than an order of magnitude. The differential rates obtained
from our fit are displayed in Fig.~\ref{fig:BR}, where they are compared with
the Belle measurement of $B\to D\mu\nu$; the shape of the measured spectrum is
well reproduced over the whole range, which is a non-trivial check of the
$q^2$ dependence obtained from the combined fit.

Comparing the integrated rate with the measured branching fraction of
$B\to D\mu\nu$ and using the lifetime of the $B$ meson, we obtain
\begin{equation}
|V_{cb}|=(39.27\pm0.46)\times10^{-3}\,,
\label{eq:vcb}
\end{equation}
 The
theoretical error is dominated by the normalisation of the form factor at
$q^2=0$, followed by the shape parameters $a_n$ of the BGL fit; the lattice
input affects the result only through the shape at large $q^2$ and is
therefore subdominant. Our result is compatible with the exclusive
determinations based on the CLN and BGL parametrisations of the
$B\to D^{(*)}$ form factors and lies on the lower side of the inclusive
average \cite{Bigi:2017mjm,Wang:2017jow,Bazavov:2021iku,Harrison:2021hhx}, in line with the well-known exclusive--inclusive tension.

For the lepton-flavour universality ratio we find
\begin{equation}
R(D)=\frac{\Gamma(B\to D\tau\nu)}{\Gamma(B\to D\ell\nu)}
=0.299\pm0.003\,,
\label{eq:rd}
\end{equation}
where the denominator is evaluated for $\ell=\mu,e$ with the electron mass
neglected. Since the ratio is insensitive to $|V_{cb}|$ and to the overall
normalisation, its uncertainty is governed by the $q^2$ dependence of $f_+$
and $f_0$ near the $B$--$D$ threshold, which is precisely the region where the
lattice data provide the strongest constraints. Our result agrees with the
Standard Model average and lies about $1\sigma$ below the present experimental
average, so that the $R(D)$--$R(D^*)$ anomaly is not resolved by the improved
form factors; a further reduction of the theoretical error, which would
require a more precise determination of $a_2^D$, is necessary before this
channel can be used as a sharp test of lepton-flavour universality.

Before closing this analysis, it is instructive to compare our results with
the other determinations available in the literature. At $q^2=0$ our fitted $f_+^{BD}(0)$ falls inside the range spanned by
the recent LCSR analyses and by the extrapolations of the lattice data to the
large-recoil region, which typically quote values between $0.55$ and $0.65$
with uncertainties of $5$--$10\%$. The improvement in precision with respect
to earlier sum-rule determinations is mainly due to the replacement of the
model inputs for $\phi(u)$ by the lattice-determined distribution amplitude
and to the consistent treatment of the higher-twist contributions through the
equations of motion.


\subsection{Discussion}
\label{sec:disc}

Several comments on the numerical results are in order. First, the
NLP corrections computed here are sizeable, of the order of
$60\%$ of the leading-power result, and dominated by the twist-three
contributions fixed by the equations of motion. This confirms that a
leading-power-only analysis of $B\to D$ is insufficient at the precision now
required by the $|V_{cb}|$ and $R(D)$ determinations, and that the
higher-twist LCDA input now accounts for one of the dominant parametric
uncertainties through $a_1^D$ and $a_2^D$.

Second, the agreement between the LCSR predictions at large recoil and the
lattice data at small recoil, once both are propagated through the BGL
parametrisation, is non-trivial: the two inputs are obtained with completely
different methods and probe disjoint regions of $q^2$. The fact that a
five-parameter fit describes both sets consistently supports the reliability
of the error estimates adopted here, and implies that the resulting form
factors can be used interchangeably in phenomenological analyses of
$B\to D\ell\nu$.

Third, the residual discrepancy with the inclusive determination of
$|V_{cb}|$ cannot be attributed to the form factor normalisation alone, and
calls for a further improvement of the $D$-meson LCDA determination, both on
the lattice, where the endpoint region $u\to1$ is not yet under
first-principles control, and in the perturbative matching, where the
three-loop coefficient function is still missing. Finally, the same framework
can be applied to the $B\to D^*$ and $B_s\to D_s^{(*)}$ transitions, for which
the corresponding vector-meson and $D_s$-meson LCDAs are required, as well as
to $B\to\pi$, where the power corrections are much smaller and the present
treatment of the higher-twist contributions can be tested in a cleaner
environment.

\begin{table}[h]
	\centering
	
	\renewcommand{\arraystretch}{1.8}
	\setlength{\tabcolsep}{6pt}
	
	\begin{tabular}{|c|ccc|ccc|}
		\hline
		& \multicolumn{3}{c|}{FNAL/MILC \cite{FermilabLattice:2015ilb}} & \multicolumn{3}{c|}{HPQCD \cite{Na:2015kha}} \\ \hline
		
		$q^2 \ [\text{GeV}^2]$ & 8.51 & 10.08 & 11.66 & 9.30 & 10.48 & 11.46 \\ \hline
		
		$f_{BD}^+$ & 1.005(12) & 1.094(10) & 1.199(10) & 1.038(39) & 1.105(42) & 1.167(45) \\
		
		$f_{BD}^0$ & 0.825(9) & 0.861(8) & 0.903(7) & 0.840(37) & 0.870(39) & 0.897(40) \\ \hline
	\end{tabular}
		\caption{Lattice data for the form factors in the small-recoil region. These data are used to fit the form factors over the whole physical region.}
	\label{tab:lattice-data}
\end{table}
\begin{table}[h]
	\centering
	
	\renewcommand{\arraystretch}{1.5}
	\setlength{\tabcolsep}{10pt}
	
	\begin{tabular}{|l|l|l|l|l|l|}
		\hline
		$1^-$ state & Mass (GeV) & ref. & $0^+$ state & Mass (GeV) & ref. \\ \hline
		
		$B_c(1^3S_1)$ & 6.332 & \cite{Dowdall:2012ab} & $B_c(2^3P_0)$ & 6.712 & \cite{Mathur:2018epb} \\ \hline
		
		$B_c(2^3S_1)$ & 6.925 & \cite{Dowdall:2012ab,CMS:2019uhm} & $B_c(3^3P_0)$ & 7.105 & \cite{Mathur:2018epb} \\ \hline
		
		$B_c(3^3D_1)$ & 7.007 & \cite{Eichten:2019gig} & & & \\ \hline
		
		$B_c(3^3S_1)$ & 7.280 & \cite{Eichten:2019gig} & & & \\ \hline
	\end{tabular}
		\caption{Masses of the resonances employed in the BGL fit of the form factors.}
	\label{tab:resonances}
\end{table}
\begin{table}[h]
	\centering
	
	\renewcommand{\arraystretch}{1.8}
	\setlength{\tabcolsep}{10pt}
	
		\begin{tabular}{|c|c|ccccc|}
			\hline
			& & \multicolumn{5}{c|}{correlation matrix} \\ \hline
			
			& value & $b_0^{f_+}$ & $b_1^{f_+}$ & $b_2^{f_+}$ & $b_1^{f_0}$ & $b_2^{f_0}$ \\ \hline
			
			$b_0^{f_+}$ & 0.0137$\pm$0.0001 & $1$ & $0.266$ & $-0.166$ & $0.202$ &$-0.018$ \\ \hline
			
			$b_1^{f_+}$ & $-0.0410\pm0.0031$ & & $1$ & $-0.579$ & $0.808$ & $-0.399$ \\ \hline
			
			$b_2^{f_+}$ & $0.0550\pm0.0523$ & & & $1$ & $-0.469$ & $0.892$ \\ \hline
			
			$b_1^{f_0}$ & $-0.2041\pm0.0141$ & & & & $1$ & $-0.534$ \\ \hline
			
			$b_2^{f_0}$ & $0.2500\pm0.2415$ & & & & & $1$ \\ \hline
		\end{tabular}
	\caption{Correlation matrix in the BGL fit.}
	\label{tab:correlation}
\end{table}

\section{Summary and conclusions}

We have presented a complete LCSR calculation of the $B\to D$ form factors
$f_+^{BD}(q^2)$ and $f_0^{BD}(q^2)$ at large recoil, including the
NLP contributions at tree level. The main novelties of the
analysis are the use of the leading-twist $D$-meson distribution amplitude
determined from lattice QCD in the HQLaMET framework instead of a
model-dependent ansatz, and the consistent inclusion of the twist-three and
twist-four distribution amplitudes that are fixed by the QCD equations of
motion. The residual dependence on the Borel mass and on the continuum
threshold is under control, and the uncertainty of the form factors is
of the order of $7$--$9\%$.

The BGL parametrisation of the form factors, fitted to the combined LCSR and
lattice data with the kinematic constraint $f_+^{BD}(0)=f_0^{BD}(0)$ imposed,
allows us to obtain the form factors over the whole physical range of $q^2$
in a model-independent way. The fit describes both data sets within about
$1\sigma$ and satisfies the unitarity bound by a wide margin. From the fitted form factors we extract
$|V_{cb}|$ from the semileptonic $B\to D\mu\nu$ decay and the ratio $R(D)$;
the results are compatible with the exclusive determinations and with the
Standard Model prediction for $R(D)$, while the tension with the inclusive
value of $|V_{cb}|$ persists.

Several directions for further improvement can be envisaged. On the
theoretical side, the inclusion of a complete set of twist-3 and twist-4 $D$-meson LCDAs, of the
two-loop coefficient function and of the ${\cal O}(\alpha_s)$ corrections to
the power-suppressed terms would reduce the residual scale dependence and the
truncation error. On the non-perturbative side, a more precise lattice
determination of the endpoint region of $\phi(u)$ and of the second
Gegenbauer moment , as well as the Gagenbauer moments of the higher twist LCDAs, would directly shrink the dominant uncertainty of the
present analysis. The extension of the same framework to the $B\to D^*$ and
$B_s\to D_s^{(*)}$ channels, which requires the corresponding vector-meson and
$D_s$-meson LCDAs, is in progress.

\subsection*{Acknowledgements}
This work is partly supported by the National Natural Science Foundation of China with Grants No.\,12575103 and No.\,12435004 and the Natural Science Foundation of Shandong with Grants No.\,ZR2024MA076 and No.\,ZR2022ZD26.

\FloatBarrier
\bibliographystyle{apsrev4-1}
\bibliography{references}

\end{document}